\documentclass[runningheads]{llncs}

\usepackage{eccv}

\usepackage{eccvabbrv}

\usepackage{graphicx}
\usepackage{booktabs}

\usepackage{indentfirst}
\usepackage{stackengine}
\usepackage{multirow}
\usepackage{graphicx}
\usepackage{amsmath}
\usepackage{amssymb}
\usepackage{booktabs}
\usepackage{tabularx}
\usepackage{wrapfig}
\usepackage{threeparttable}

\usepackage[accsupp]{axessibility}  

\usepackage{hyperref}

\usepackage{orcidlink}

\begin{document}

\title{BurstM: Deep Burst Multi-scale SR using Fourier Space with Optical Flow} 

\titlerunning{BurstM}

\author{
EungGu Kang$^1$\qquad Byeonghun Lee$^2$\qquad Sunghoon Im$^1$\thanks{Co-corresponding authors.}\qquad Kyong Hwan Jin$^{2*}$}

\authorrunning{EG Kang et al.}

\institute{Daegu Gyeongbuk Institute of Science and Technology (DGIST), Korea \and
Korea University, Korea\\
\email{eunggukang@gmail.com, sunghoonim@dgist.ac.kr, \{byeonghun\textunderscore lee, kyong\textunderscore jin\}@korea.ac.kr}}

\maketitle

\vspace{-0.3in} 
\begin{abstract}
  Multi frame super-resolution (MFSR) achieves higher performance than single image super-resolution (SISR), because MFSR leverages abundant information from multiple frames. Recent MFSR approaches adapt the deformable convolution network (DCN) to align the frames. However, the existing MFSR suffers from misalignments between the reference and source frames due to the limitations of DCN, such as small receptive fields and the predefined number of kernels. From these problems, existing MFSR approaches struggle to represent high-frequency information.
   To this end, we propose Deep Burst Multi-scale SR using Fourier Space with Optical Flow (BurstM). The proposed method estimates the optical flow offset for accurate alignment and predicts the continuous Fourier coefficient of each frame for representing high-frequency textures. In addition, we have enhanced the network's flexibility by supporting various super-resolution (SR) scale factors with the unimodel. We demonstrate that our method has the highest performance and flexibility than the existing MFSR methods. Our source code is available at \url{https://github.com/Egkang-Luis/burstm}
    \keywords{Super resolution \and Burst super resolution \and Implicit neural representation \and Fourier features}
    \vspace{-10pt}
\end{abstract}

\section{Introduction}
\vspace{-0.05in} 

    Single image super-resolution (SISR) is the major ill-posed problem in low-level computer vision. SISR aims to enhance the image quality of low-resolution by estimating sub-pixel information. After the success of deep learning, the most SISR~\cite{cnn_base_approaches_one,cnn_base_approaches_two,cnn_base_approaches_three,cnn_base_approaches_four,cnn_base_approaches_five} adapt the learning-based method to estimate high-resolution images. However, SISR has a limitation in high-frequency representation of the capacity due to a lack of information.
    
    Super-resolution researchers focus on multi-frame super-resolution (MFSR) due to the accurate representation of high-frequency textures. Multiple frames provide additional sub-pixel priors, and MFSR utilizes rich information with alignment between the reference and source frames. 
    \begin{figure}[ht!]
        \centering
        \includegraphics[width=0.7\linewidth]{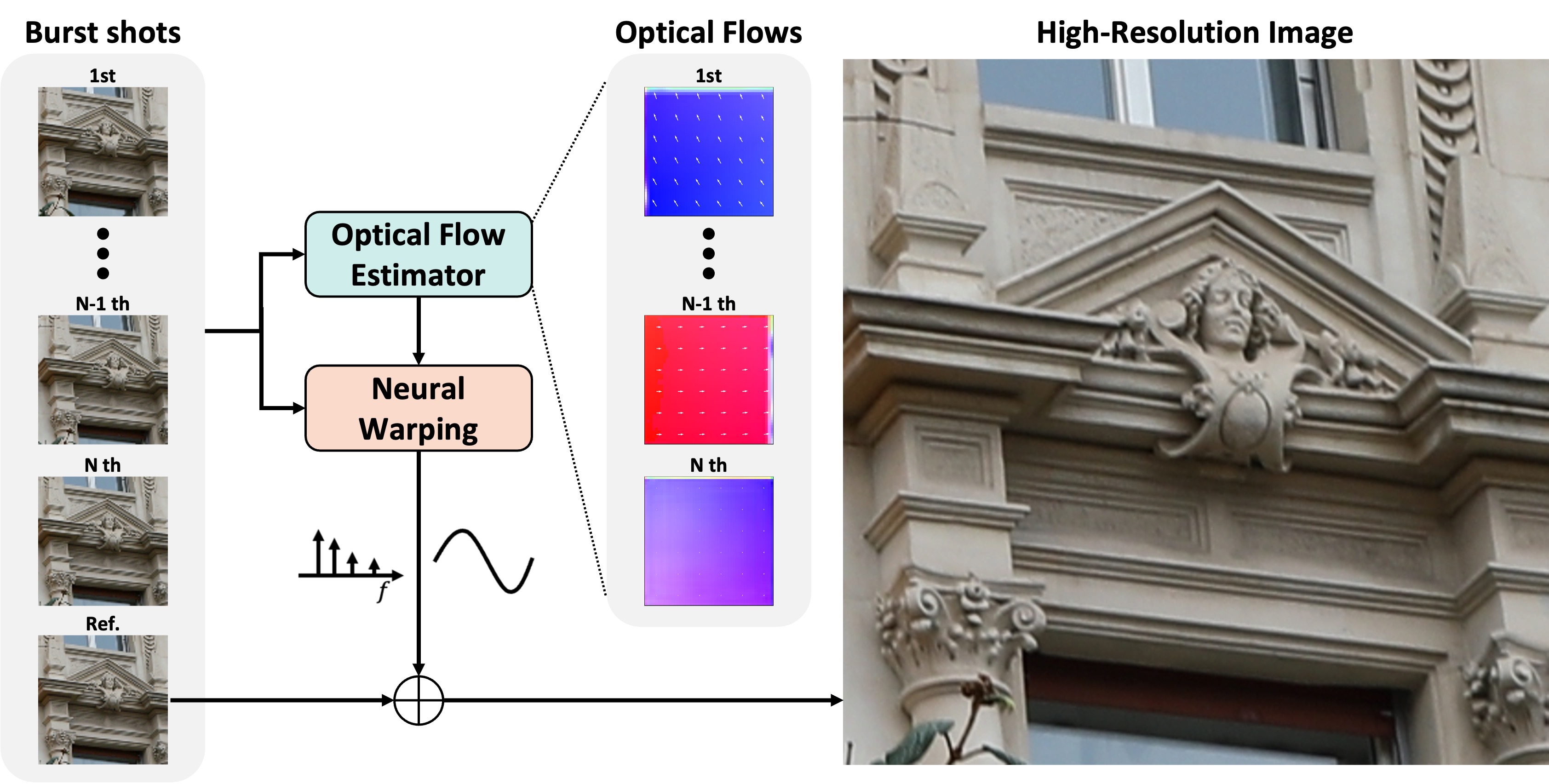}
        \caption{\textbf{Overview of proposed Burst SR network.} The proposed method (BurstM) predicts high-resolution images from multiple low-resolution images. The neural warping performs high-precision warping with upscaling on Fourier space. Additionally, the neural warping contributes to the prediction of an accurate high-resolution images.}
        
        \label{fig:simple_pipeline}
        \vspace{-15pt}
    \end{figure}
    However, multiple frames take similar scenes with degradations such as noise and object movement. MFSR requires the accurate alignment performance to utilize low-quality frames, because the inaccurate alignment performance can lead to the blur artifacts in the prediction result. Thus, the alignment process is the critical challenge of MFSR. 
    
    Like SISR, the latest MFSR~\cite{dbsr_burstsr_dataset, mfir, bipnet, gmtnet, burstormer, bsrt} employs a deep learning method. They adapt similar architectures, which is complicated feature extraction methods using CNN or transformer~\cite{liu2021Swin,liang2021swinir,DBLP:conf/cvpr/Chen000DLMX0021} and alignment using DCN~\cite{deformable_convolution_network} to reconstruct the high-resolution images. Recent MFSR~\cite{bipnet, gmtnet, burstormer, bsrt} adapts multi-layer DCN structures to extend a small receptive field of DCN, but even this approach still has limited receptive fields. In addition, the predefined number of DCN kernels restricts representing performance during aligning features. The fixed SR scale is another limitation of existing MFSR due to their upsampling method. Most networks~\cite{bipnet, gmtnet, burstormer, bsrt} generally adapt pixel shuffle (PS)~\cite{pixel_shuffle} as an upsampling module due to the effectiveness of PS. Nevertheless, PS restricts the network's flexibility, e.g., $\times4$ SR, because PS can conduct upsampling by a predetermined scale, which is the square of the channel number. If the network predicts a different SR scale, it must re-train with changing architecture or additional downsampling after a predefined SR scale.

    To this end, we propose Deep Burst Multi-scale SR using Fourier Space with Optical Flow (BurstM), which has a high representation capability and flexible scale SR ($\times2, \times3, \times4$) with computational efficiency. BurstM adapts Optical Flow~\cite{optical_flow} as a solution for mitigating the limitations of DCN. Estimated optical flow provides correlated offsets between reference and source frames. In order to represent high-frequency textures using well-aligned information, our network estimates Fourier information. To address the fixed scale factor of MFSR, we employ INR~\cite{inr} inspired by recent successful researches~\cite{nis, lte}. 
  
    To summarize, our contribution is as follows:
    \vspace{-5pt}
    \begin{itemize}
    \setlength{\leftmargin}{-.35in}
      \item Our network conducts precise alignment using optical flow without DCN on the Fourier space.
      \item We solve the fixed-scale issue in MFSR and demonstrate our  flexible networks of multi-scale SR.
      \item Extensive experiments demonstrate that our approach outperforms existing approaches in both image quality and computational efficiency.
    \end{itemize}
    \vspace{5pt}
    
\section{Related Work}
    \vspace{-5pt}
    \noindent{\textbf{Multi Frame Super-Resolution (MFSR)}}
    The first MFSR from Tsai~\cite{tsai1984multiframe} restores high-resolution images in the frequency domain. The following approaches
    ~\cite{ peleg1987improving,irani1991improving,bascle1996motion,elad1997restoration,hardie1998high} have been introduced using the spatial domain and the maximum posterior. Deep learning-based MFSR approaches~\cite{dbsr_burstsr_dataset,mfir,bipnet,gmtnet,burstormer, bsrt} have arisen due to higher performance than the conventional methods. Furthermore, since the introduction of DCN~\cite{deformable_convolution_network} as the effective alignment method, many approaches~\cite{bipnet, zhang2022self,burstormer,gmtnet, bsrt} have adopted DCN because a biased position of DCN kernel contributes to outstanding alignment performance. However, the small kernel size and a fixed number of kernels restrict the alignment performance. A small kernel size ($3\times3$) of DCN has a small receptive field, which makes it impossible to cover global alignments. To address it, recent methods~\cite{bipnet,gmtnet, burstormer, bsrt} employ multi-layer DCN in varying spatial information, but it is not enough. Another issue is the fixed number of DCN kernels. Even if DCN cannot find the relevant information, DCN still needs to use the number of predefined kernels, leading to blurry and noisy images. Our BurstM employs optical flow to acquire a broader receptive field and utilize only one-to-one correlated information to tackle this issue.\\
    
    \noindent{\textbf{Optical Flow}}
    Optical flow is a popular method to describe the pixel-level correspondence of object displacements between frames. The conventional methods~\cite{optical_flow, hirschmuller2007stereo, triggs2000bundle} address the data fidelity and the regularization term as an energy optimization problem. However, they often struggle to estimate accurate results in various scenarios, such as substantial object displacements or small-sized objects. Recent learning-based methods~\cite{sun2018pwc, ranjan2017optical, teed2020raft} outperform the conventional method's accuracy and computational cost using various structures such as hierarchical design and progressive estimation. Many tasks, such as video super-resolution~\cite{chan2021basicvsr, lai2017deep}, action recognition~\cite{simonyan2014two}, and autonomous driving~\cite{janai2020computer}, predict the optical flow to find pixel correspondences. However, MFSR task does not effectively utilize optical flow due to computational cost. On the other hand, the proposed network (BurstM) has a simple structure and lightweights, allowing us to incorporate an optical flow estimator into the network.\\
    
    \noindent{\textbf{Implicit Neural Representation (INR)}}
    One of the popular representation methods for the continuous-domain signal is INR~\cite{inr}. Neural networks designed for INR learn and represent complex structures implicitly. The various tasks employ INR in image processes such as 2D and 3D shapes. Despite such success, INR parameterized by a standalone MLP with the ReLU has a spectral bias problem~\cite{spectral_bias_one, spectral_bias_two, spectral_bias_three} that fails to represent high-frequency details known as a spectral bias. Recent approaches~\cite{lte, chen2021learning, nis} tackle a spectral bias through positional encoding and Fourier feature mapping, demonstrating a flexible SR network. We have adopted INR method to utilize the flexible characteristic.

\section{Problem Formulation}
\begin{figure*}[t!]
        \centering
        \includegraphics[width=1\linewidth]{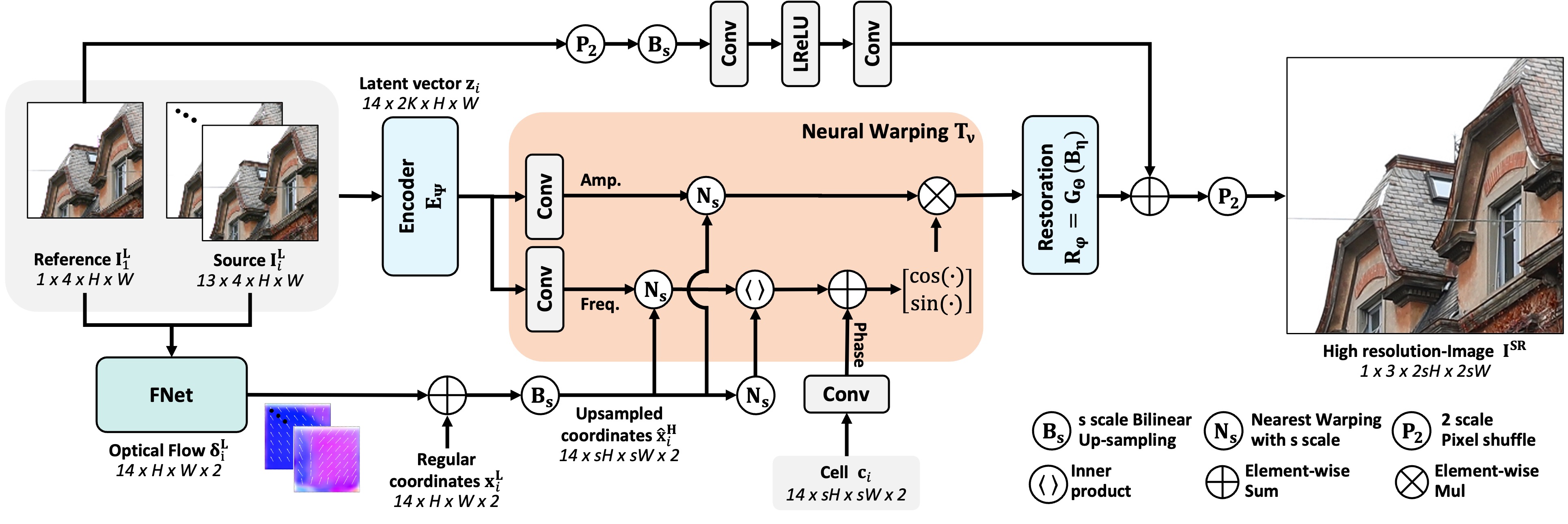}
        \caption{\textbf{Proposed BurstM Network}. The network estimates optical flow between the reference and source frames using FNet. The neural warping $\mathbf{T_{\boldsymbol{\nu}}}$ emphasizes the high-frequency details and implements the warping on Fourier space. Reconstruction module $\mathbf{R_{\boldsymbol{\varphi}}}$ includes the blender $\mathbf{B_{\boldsymbol{\eta}}}$ and the decoder $\mathbf{G_{\boldsymbol{\Theta}}}$. The blender $\mathbf{B_{\boldsymbol{\eta}}}$ matches color differences and merges multiple frames into a single frame. The Decoder $\mathbf{G_{\boldsymbol{\Theta}}}$ predicts the final output. The gray color indicates a single operation such a convolution and LeakyReLU activation and the others include nonlinear activation}
        \label{fig:Overall_architecture}
        \vspace{-10pt}
    \end{figure*}

    \vspace{-0.05in}
    Our purpose is to reconstruct the high-resolution image $\mathbf{I}^{\mathbf{SR}}\in\mathbb{R}^{3 \times 2sH \times 2sW}$ with multiple scale factors s from the set of a low-resolution images $\{\mathbf{I}^\mathbf{L}_{i}\}_{i=1}^N$, where $\mathbf{I}^\mathbf{L}_{i}\in\mathbb{R}^{4 \times H\times W}$, $N$ is the number of frame and $H, W$ are the spatial sizes of low resolution image, and 4-channels are due to `RGGB' domain. The dimension of a high-resolution image are increased by a factor of 2s, because the domain is changed from 4-channels `RGGB' to 3-channels `sRGB' in the network (in Figure \ref{fig:Overall_architecture}). 
    \vspace{-10pt}
    
\subsection{Optical Flow Estimator}
We adopt FNet (optical flow estimator) of FRVSR~\cite{frvsr}. FNet predicts optical flow on low-resolution grid as below:
\begin{equation}
\boldsymbol{\delta}^\mathbf{L}_{i}=FNet\mathbf{(I}^\mathbf{L}_{i}),  \quad i=1,\ldots,N,
\end{equation}
where $\boldsymbol{\delta}^\mathbf{L}_{i} \in \mathbb{R}^{H \times W \times 2}$ is the set of optical flow offsets between a reference and source frames on low-resolution grid. Our model adds offsets $\boldsymbol{\delta}^\mathbf{L}_{i}$ into regular coordinates $\mathbf{x}^\mathbf{L}_{i} \in [-1, 1]^{{H \times W \times 2}}$ and applies s scale upsampling.
\begin{equation}
    \mathbf{\hat{x}}^\mathbf{H}_{i}=\mathbf{B_s}(\boldsymbol{\delta}^\mathbf{L}_{i}+\mathbf{x}^\mathbf{L}_i), \quad i=1,\ldots,N,
\end{equation}
where $\mathbf{B_s(\cdot)}$ is bilinear upsampling operator with a scale factor of s, and $\mathbf{\hat{x}}^\mathbf{H}_{i} \in \mathbb{R}^{sH \times sW \times 2}$ is the upscaled coordinate with an offset defined by scale factor s.
\vspace{-10pt}

\subsection{Learnable Neural Warping}
\vspace{-5pt}
Our encoder ($\mathbf{E}_{\boldsymbol{\psi}}(\cdot) : \mathbb{R}^{4} \mapsto \mathbb{R}^{2K})$, which is a learnable super-resolution (SR) backbone~\cite{cnn_base_approaches_two} without upsampling, extracts latent vectors $\mathbf{z}_{i}$. NW ($\mathbf{T}_{\boldsymbol{\nu}}$) takes the latent vectors $\mathbf{z}_{i}$ for warping based on a Fourier information. NW consists of three components: an amplitude estimator ($\mathbf{g}_\mathbf{a}(\cdot) : \mathbb{R}^{2K} \mapsto \mathbb{R}^{2K}$), a frequency estimator ($\mathbf{g}_\mathbf{f}(\cdot) : \mathbb{R}^{2K} \mapsto \mathbb{R}^{2K}$) and, a phase estimator ($\mathbf{g}_\mathbf{p}(\cdot) : \mathbb{R}^2 \mapsto \mathbb{R}^{K}$). $2K$ denotes the number of channels in the latent vector. Amplitude and frequency estimators utilize the latent vector $\mathbf{z}_{i}$ to estimate the Fourier coefficients. After that, the nearest interpolation $\mathbf{N}_\mathbf{s}$ is performed using estimated coordinates $\mathbf{\hat{x}}^\mathbf{{L}}_{i}$. The input of a phase estimator is a cell $\mathbf{c}_{i}\in \mathbb{R}^{sH \times sW \times 2}$, which is consist of the divisor of the input frame’s spatial sizes ($1/H, 1/W$). As in ~\cite{chen2021learning,Local_Implicit_Grid_CVPR20}, we use relative coordinates ($\boldsymbol{\delta}_{i} = \mathbf{\hat{x}}^\mathbf{H}_i - \mathbf{x}^\mathbf{H}_{i}$) known as a local grid to predict the continuous signal where $\mathbf{\hat{x}}^\mathbf{H}_{i} \in \mathbb{R}^{sH \times sW \times 2}$. The designed formulation is as follows:
\vspace{-10pt}

\begin{gather}
    \mathbf{z}_{i}=\mathbf{E_{\boldsymbol{\psi}}}(\mathbf{I}^\mathbf{L}_{i}), \\
    \mathbf{F}_{i}=\mathbf{N_s}(\mathbf{g}_\mathbf{f}(\mathbf{z}_{i}), \mathbf{\hat{x}}^\mathbf{H}_{i}), \quad 
    \mathbf{A}_{i}=\mathbf{N_s}(\mathbf{g}_\mathbf{a}(\mathbf{z}_{i}), \mathbf{\hat{x}}^\mathbf{H}_{i}), \quad 
    \mathbf{ph}_{i}=\mathbf{g}_\mathbf{p}(\mathbf{c}_{i}),\\
    \mathbf{m}_{i} = \mathbf{A}_{i}\odot
    \begin{bmatrix}
        \mathbf{\cos}(\pi (\mathbf{\langle F}_{i}, \boldsymbol{\delta}_{i}\rangle + \mathbf{ph}_{i})\\
        \mathbf{\sin}(\pi (\mathbf{\langle F}_{i}, \boldsymbol{\delta}_{i}\rangle + \mathbf{ph}_{i})
    \end{bmatrix},\\
    \text{where}, \quad i=1,\ldots,N,\quad \mathbf{{\boldsymbol{\nu}}=[f;a;p]}. \nonumber
\end{gather}
\vspace{-30pt}

\subsection{Reconstruction Module}
\vspace{-5pt}
To reconstruct the high-resolution image using estimated features in Fourier space, we employ the reconstruction module $\mathbf{R_{\boldsymbol{\varphi}}}$; Blender ($\mathbf{B_{\boldsymbol{\eta}}(\cdot)} : \mathbb{R}^{N \times 2K} \mapsto \mathbb{R}^{1 \times 2K}$) and MLP decoder ($\mathbf{G_\Theta(\cdot)} : \mathbb{R}^{2K} \mapsto \mathbb{R}^{12}$).
\begin{equation}
    \mathbf{h} = \mathbf{R}_{\boldsymbol{\varphi}}(\{\mathbf{m}_{i}\}^{N}_{i=1}) = \mathbf{G}_{\mathbf{\Theta}}(\mathbf{B}_{\boldsymbol{\eta}}(\{\mathbf{m}_{i}\}^{N}_{i=1})),
\end{equation}
where $\mathbf{h}\in\mathbb{R}^{1 \times 12\times sH\times sW}$ is the merged Fourier features of burst shots. The blender $\mathbf{B_{\boldsymbol{\eta}}}$, which is the same module with the encoder $\mathbf{E_{\boldsymbol{\psi}}}$, reduces color differences between frames and merges multiple frames. MLP decoder ($\mathbf{G_{\Theta}}$) estimates `sRGB' information.
\vspace{-10pt}

\subsection{Skip connection with upsampling}
\vspace{-5pt}
To prevent learning on the low-frequency only, we utilize the upscaled skip connection $(\mathbf{Q}_{\boldsymbol{\xi}}(\cdot) : \mathbb{R}^{4 \times H \times W} \mapsto \mathbb{R}^{12 \times sH \times sW})$ about the reference frame $\mathbf{I}^{\mathbf{L}}_{1}$. Finally, the spatial size is increased using pixel shuffle ($\mathbf{P_2}(\cdot) : \mathbb{R}^{12 \times sH \times sW} \mapsto \mathbb{R}^{3 \times 2sH \times 2sW}$). The proposed BurstM can be formulated as follow:
\begin{gather}
    \mathbf{I}^{\mathbf{SR}}=\mathbf{P}_{\mathbf{2}}(\mathbf{h}+\mathbf{Q}_{\boldsymbol{\xi}}(\mathbf{I}^{\mathbf{L}}_{1}))
\end{gather}
\vspace{-30pt}

\subsection{Loss function}
\vspace{-5pt}
We use two loss functions for both SyntheticBurst~\cite{synthetic_datset} and BurstSR~\cite{dbsr_burstsr_dataset}. The MFSR datasets~\cite{dbsr_burstsr_dataset,synthetic_datset} does not provide the ground truth of the optical flow map between frames. Thus, we use L2 loss term to use the photometric loss.
    \begin{equation}
    \mathcal{L_{OF}}=\sum_{i=2}^N \left\|{Warp(\mathbf{I}^\mathbf{L}_{i}, \boldsymbol{\delta}^\mathbf{L}_{i})-\mathbf{I}^\mathbf{L}_1}\right\|_2 =\sum_{i=2}^N ({Warp(\mathbf{I}^\mathbf{L}_{i}, \boldsymbol{\delta}^\mathbf{L}_{i})-\mathbf{I}^\mathbf{L}_1})^{2},
    \end{equation}
    where $Warp(\mathbf{\cdot,\cdot})$ is the bilinear warping operator using optical flow from FNet, and $\mathbf{I}^\mathbf{L}_{1}$ is the reference frame.\\
    
    \noindent{\textbf{SyntheticBurst dataset}} Input frames are generated from the ground truth. Thus, we update our model using L1 loss between predicted image and the ground truth.
    \begin{equation}
        \mathcal{L_{SR}}= \left\|\mathbf{I^{SR}}-\mathbf{I^{GT}}\right\|_1.
    \end{equation}
    Thus, the total loss function during training is as follows:
    \begin{equation}
        \mathcal{L}_{Synthetic}=\mathcal{L_{SR}}+\mathcal{L_{OF}.}
    \end{equation}
    \textbf{BurstSR dataset} We employs an aligned L1 loss for the BurstSR~\cite{dbsr_burstsr_dataset} dataset due to mismatching between low-resolution images and a ground truth image. An aligned L1 loss utilizes pre-trained weight of PWCNet~\cite{sun2018pwc} as follows:
    \begin{gather}
        \boldsymbol{\delta^{A}}= PWCNet(\mathbf{I^{SR}, I^{GT}),}\\
        \mathcal{L}_{AlignedSR}= \left\|Warp\mathbf{(\mathbf{I^{SR}}, \boldsymbol{\delta^{A}})-\mathbf{I}^{GT}}\right\|_1,
    \end{gather}

\noindent{where $\boldsymbol{\delta}^\mathbf{A}$ is optical flow between the predicted image on high-resolution grid and ground truth using PWCNet~\cite{sun2018pwc}. Total loss function for BurstSR~\cite{dbsr_burstsr_dataset} dataset is as follows:}
\begin{equation}
    \mathcal{L}_{Real}=\mathcal{L}_{AlignedSR}+\mathcal{L_{OF}.}
\end{equation}

\section{Experiment}
    \subsection{Network details}
    \noindent\textbf{Optical Flow Estimator} (\textit{FNet}) Our network adapts FNet from FRVSR~\cite{frvsr}. To apply FNet into BurstSR task, we modify the stem of FNet from `sRGB' to `RGGB' images.\\
    
    \noindent\textbf{Learnable Neural Warping ($\mathbf{T_{\boldsymbol{\nu}}}$)} NW estimates the amplitude, frequency, and phase from burst shots with a single convolution operation to represent the Fourier information. Amplitude and frequency estimators have 128 channels of the convolutions with $3\times3$ kernel, phase estimator has 64 channels of the convolutions with $1\times1$ kernel.\\
    
    \noindent\textbf{Encoder ($\mathbf{E_{\boldsymbol{\psi}}}$) and Reconstruction module ($\mathbf{R_{\boldsymbol{\varphi}}}$)} The reconstruction module consists of Blender $\mathbf{B_{\boldsymbol{\eta}}}$ and MLP decoder $\mathbf{G_\Theta}$. Both Encoder $\mathbf{E_{\boldsymbol{\psi}}}$ and Blender $\mathbf{B_{\boldsymbol{\eta}}}$ are EDSR~\cite{cnn_base_approaches_two} without upsampling module and 128 channels. MLP Decoder , $\mathbf{G_\Theta}$, is organized by 4-layer of $1\times1$ convolution with 256 channels and ReLU activation. 
    
    \subsection{Implementation details}
     BurstM is an end-to-end trainable network and is possible to train without pre-trained weights. We apply an AdamW optimizer~\cite{adamw} with the Cosine annealing strategy~\cite{cosine_annealing} from $10^{-4}$ to $10^{-6}$ learning rate. For SyntheticBurst~\cite{synthetic_datset}, we train BurstM with L1 loss for the output image and L2 loss for optical flows. For BurstSR~\cite{dbsr_burstsr_dataset}, We fine-tune the pre-trained BurstM of SyntheticBurst with L2 loss for optical flow and aligned L1 loss~\cite{dbsr_burstsr_dataset} for the output image. Aligned L1 loss utilizes PWCNet~\cite{sun2018pwc} for the output image. The reason for the usage of the aligned L1 Loss is that there are no LR images aligned with the GT image because they were captured using different cameras. We train our model using four RTX3090 GPUs and training times are 10 days for SyntheticBurst and 8 hours for BurstSR. The summary of the training strategy is in Table \ref{tab:training_strateges}.
    
    \begin{table}[ht]
    \vspace{-20pt}
    \centering
    \begin{tabular}{c|cccc}
    \toprule
                   & SyntheticBurst~\cite{synthetic_datset} & BurstSR~\cite{dbsr_burstsr_dataset} \\ \hline\hline
    Training Epoch & 300                                   & 25                                 \\
    Optimizer      & \multicolumn{2}{c}{ADAMW~\cite{adamw}}                                      \\
    Learning rate  & \multicolumn{2}{c}{$10^{-4}\sim 10^{-6}$}                                  \\
    LR scheduler   & \multicolumn{2}{c}{Cosine annealing~\cite{cosine_annealing}}                \\
    Batch sizes    & \multicolumn{2}{c}{4}                                                      \\
    Main Loss      & L1                                    & Aligned L1                         \\
    Photo metric Loss & \multicolumn{2}{c}{L2}                                                  \\
    \multirow{3}{*}{Input sizes}    & $\times2 : 96\times96$ & \multirow{3}{*}{$\times4 : 80\times80$}     \\
                                    & $\times3 : 64\times64$ &                                  \\
                                    & $\times4 : 48\times48$ &                                  \\
    
    Training time  & 10 days                               & 8 hours                            \\ \hline
    \end{tabular}
    
    \caption{\textbf{Training Strategies}. Training epoch and main loss, input sizes are different. Except for these factors, we use same training strategies for both SyntheticBurst~\cite{synthetic_datset} for BurstSR~\cite{dbsr_burstsr_dataset} datasets.}
    \vspace{-20pt}
    \label{tab:training_strateges}
    \end{table}
    
    \noindent\textbf{SyntheticBurst Dataset~\cite{synthetic_datset}} The dataset contains burst frames which are 46,839 set for training and 300 set for validation. Each burst set contains 14 low-resolution burst frames, which are generated from `sRGB' as follows. We utilized the unprocessing~\cite{inverse_camera_pipeline}. Subsequently, we apply random rotations and translations. We employ random down-sampling ($\times2$, $\times3$, $\times4$) to generate low-resolution images of multiple scales. Finally, we apply Bayer mosaicking and add random noise. We trained the proposed model with batch size 4 during 300 epochs, and the training patch size is changed per scale factors in Table \ref{tab:training_strateges}.\\
    
    \noindent\textbf{BurstSR Dataset~\cite{dbsr_burstsr_dataset}} Input and GT are captured by different cameras: Smartphone and DSLR, respectively. Both low and high resolution images have 200 raw burst sets. The low-resolution set contains 14 frames, and we extract 5,405 patches for training and 882 patches for validation. Unlike SyntehticBurst dataset~\cite{synthetic_datset}, the scale factor between low-resolution and high-resolution is fixed. Thus, we train BurstM only for $\times$4 scale factor. Nevertheless, BurstM can accurately predict high-resolution images for multiple scale factors ($\times2$, $\times3$, $\times4$), including clear representation of high-frequency textures without the boundary artifacts.
    
    \subsection{Validation Results}
    We show the qualitative comparison with the existing methods~\cite{dbsr_burstsr_dataset, mfir, bipnet, gmtnet, burstormer, bsrt} and the visual comparison with latest three methods~\cite{bipnet, burstormer, bsrt} using the pre-trained weights from authors.

    \begin{table}[h]
    \centering
        \begin{tabular}{c|c|cccccc|cc}
        \toprule
           & \multirow{3}{*}{\begin{tabular}[c]{@{}c@{}} \# Params.\\ M\end{tabular}} & \multicolumn{6}{c|}{SyntheticBurst}                                                   & \multicolumn{2}{c}{BurstSR} \\
           &                                                                      & \multicolumn{2}{c|}{x2}      & \multicolumn{2}{c|}{x3}      & \multicolumn{2}{c|}{x4} & \multicolumn{2}{c}{x4}      \\
           &                                                                      & PSNR & \multicolumn{1}{c|}{SSIM} & PSNR & \multicolumn{1}{c|}{SSIM} & PSNR        & SSIM        & PSNR          & SSIM          \\ \hline
        Bicubic          & -     & 38.30 & \multicolumn{1}{c|}{0.948}  & 33.94 & \multicolumn{1}{c|}{0.886} & 33.02 & 0.862 & 42.55 & 0.962 \\
        DBSR~\cite{dbsr_burstsr_dataset} & 13.01 & 40.51 & \multicolumn{1}{c|}{0.965}  & 40.11 & \multicolumn{1}{c|}{0.959} & 40.76 & 0.959 & 48.05 & 0.984 \\
        MFIR~\cite{mfir}                 & 12.13 & \textcolor{blue}{41.25} & \multicolumn{1}{c|}{\textcolor{blue}{0.971}}  & 41.81 & \multicolumn{1}{c|}{0.972} & 41.56 & 0.964 & 48.33 & 0.985 \\
        BIPNet$^{\dag}$~\cite{bipnet}           & 6.7   & 37.58 & \multicolumn{1}{c|}{0.928}  & 40.83 & \multicolumn{1}{c|}{0.955} & 41.93 & 0.960 & 48.49 & 0.985 \\
        GMTNet$^{\dag}$~\cite{gmtnet}           & -     & -     & \multicolumn{1}{c|}{-}  & -     & \multicolumn{1}{c|}{-} & 42.36 & 0.961 & \textcolor{blue}{48.95} & \textcolor{blue}{0.986} \\
        Burstormer$^{\dag}$~\cite{burstormer}       & 2.5   & 37.06 & \multicolumn{1}{c|}{0.925}  & 40.26 & \multicolumn{1}{c|}{0.953} & 42.83 & \textcolor{blue}{0.973} & 48.82 & \textcolor{blue}{0.986} \\
        BSRT-Small$^{\dag}$~\cite{bsrt}       & 4.92  & 40.64     & \multicolumn{1}{c|}{0.966}  & 42.30     & \multicolumn{1}{c|}{0.975} & 42.72 & 0.971 & 48.48 & 0.985 \\
        BSRT-Large$^{\dag}$~\cite{bsrt}       & 20.71 & 40.33     & \multicolumn{1}{c|}{0.965}  & \textcolor{blue}{42.87}     & \multicolumn{1}{c|}{\textcolor{blue}{0.979}} & \textcolor{red}{43.62} & \textcolor{red}{0.975} & 48.57 & \textcolor{blue}{0.986} \\ \hline
        {\bf BurstM (Ours)}    & 14.0  & \textcolor{red}{46.01} & \multicolumn{1}{c|}{\textcolor{red}{0.985}}  & \textcolor{red}{44.79} & \multicolumn{1}{c|}{\textcolor{red}{0.982}} & \textcolor{blue}{42.87} & \textcolor{blue}{0.973} & \textcolor{red}{49.12} & \textcolor{red}{0.987} \\
        \end{tabular}
        \caption{\textbf{Quantitative comparison on \underline{SyntheticBurst}} and \underline{\textbf{BurstSR}}. Quantitative comparisons of BurstSR such as PSNR and SSIM can only be measured for $\times4$, due to LR and HR being provided in fixed sizes. \textcolor{red}{Red} and \textcolor{blue}{Blue} colors indicate the best and the second performance, respectively. ${\dag}$ indicates Transformer-based method.}
        \label{tab:SR_result}
        \vspace{-0.2in}
    \end{table}

    \begin{figure}[h!]
    \centering
        \includegraphics[width=0.75in]{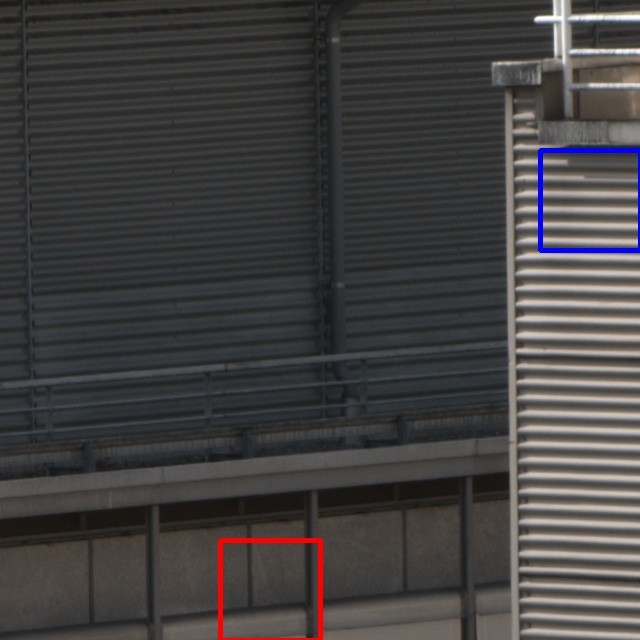}
        \includegraphics[width=0.75in]{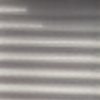}
        \includegraphics[width=0.75in]{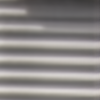}
        \includegraphics[width=0.75in]{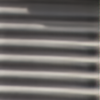}
        \includegraphics[width=0.75in]{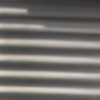}
        \includegraphics[width=0.75in]{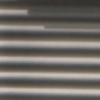}\\
        
        \includegraphics[width=0.75in]{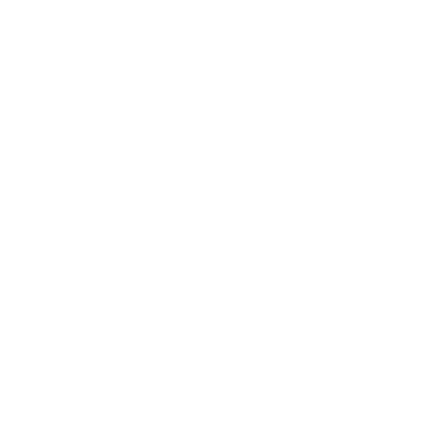}
        \includegraphics[width=0.75in]{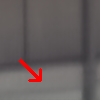}
        \includegraphics[width=0.75in]{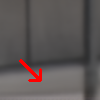}
        \includegraphics[width=0.75in]{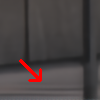}
        \includegraphics[width=0.75in]{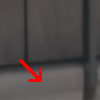}
        \includegraphics[width=0.75in]{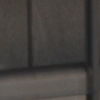}\\
        
        \includegraphics[width=0.75in]{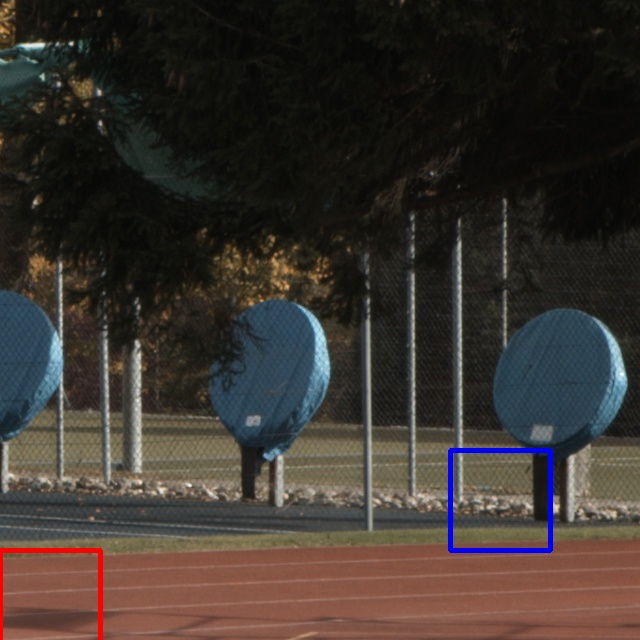}
        \includegraphics[width=0.75in]{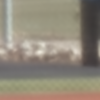}
        \includegraphics[width=0.75in]{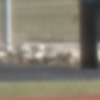}
        \includegraphics[width=0.75in]{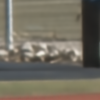}
        \includegraphics[width=0.75in]{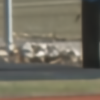}
        \includegraphics[width=0.75in]{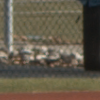}\\
        \includegraphics[width=0.75in]{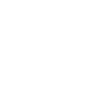}
        \includegraphics[width=0.75in]{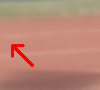}
        \includegraphics[width=0.75in]{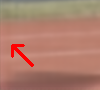}
        \includegraphics[width=0.75in]{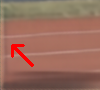}
        \includegraphics[width=0.75in]{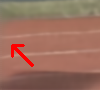}
        \includegraphics[width=0.75in]{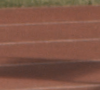}\\
        
        \includegraphics[width=0.75in]{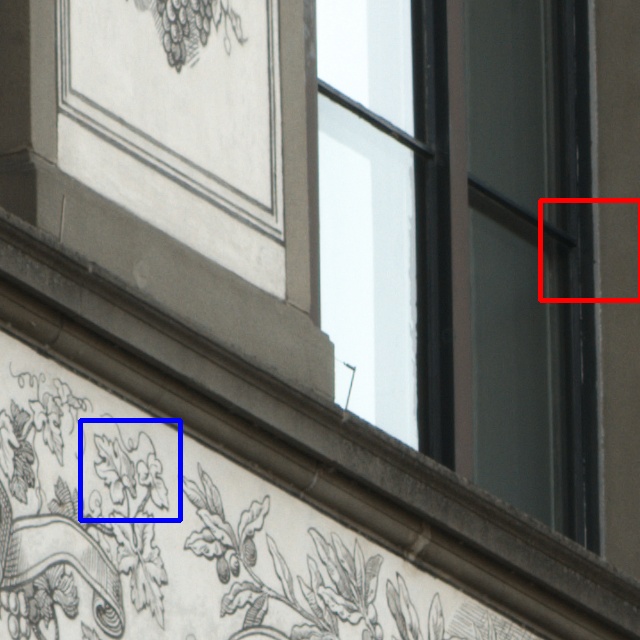}
        \includegraphics[width=0.75in]{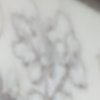}
        \includegraphics[width=0.75in]{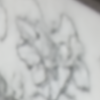}
        \includegraphics[width=0.75in]{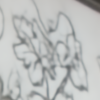}
        \includegraphics[width=0.75in]{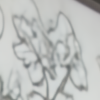}
        \includegraphics[width=0.75in]{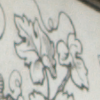}\\
        
        \stackunder[2pt]{\includegraphics[width=0.75in]{Figures/white.png}}{Full Image}
        \hspace{0.007in}
        \stackunder[2pt]{\includegraphics[width=0.75in]{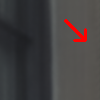}}{BIPNet}
        \stackunder[2pt]{\includegraphics[width=0.75in]{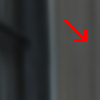}}{Burstormer}
        \stackunder[2pt]{\includegraphics[width=0.75in]{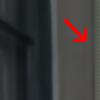}}{BSRT-L}
        \stackunder[2pt]{\includegraphics[width=0.75in]{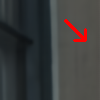}}{\textbf{BurstM}}
        \stackunder[2pt]{\includegraphics[width=0.75in]{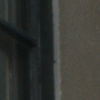}}{GT}
        \vspace*{-5pt}
        \caption{\textbf{Visual comparison for $\times$4 super-resolution for \underline{BurstSR}.} BurstM shows clearer textures and correct color than the others \underline{\textbf{without the boundary artifacts.}}}
        \label{fig:x4_comparison_for_real_burst}
        \vspace{-30pt}
    \end{figure}

\begin{figure*}[htb!]
        \footnotesize
        \centering
        \raisebox{0.01in}{\rotatebox{90}{Burstormer}}
        \includegraphics[scale=0.46]{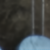}
        \includegraphics[scale=0.46]{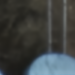}
        \includegraphics[scale=0.46]{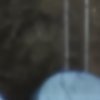}
        \includegraphics[scale=0.46]{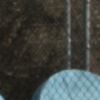}
        \includegraphics[scale=0.46]{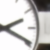}
        \includegraphics[scale=0.46]{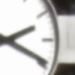}
        \includegraphics[scale=0.46]{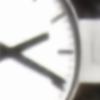}
        \includegraphics[scale=0.46]{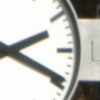}

        \raisebox{0.05in}{\rotatebox{90}{BSRT-L}}
        \includegraphics[scale=0.46]{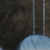}
        \includegraphics[scale=0.46]{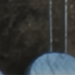}
        \includegraphics[scale=0.46]{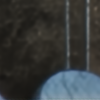}
        \includegraphics[scale=0.46]{Figures/Multi_scale_SR_on_BurstSR/0105_0013_gt.png}
        \includegraphics[scale=0.46]{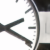}
        \includegraphics[scale=0.46]{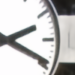}
        \includegraphics[scale=0.46]{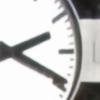}
        \includegraphics[scale=0.46]{Figures/Multi_scale_SR_on_BurstSR/0105_0009_gt.png}

        \raisebox{0.1in}{\rotatebox{90}{\textbf{Ours}}}
        \stackunder[2pt]{\includegraphics[scale=0.46]{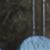}}{\textbf{$\times$2}}
        \stackunder[2pt]{\includegraphics[scale=0.46]{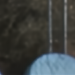}}{\textbf{$\times$3}}
        \stackunder[2pt]{\includegraphics[scale=0.46]{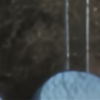}}{\textbf{$\times$4}}
        \stackunder[2pt]{\includegraphics[scale=0.46]{Figures/Multi_scale_SR_on_BurstSR/0105_0013_gt.png}}{\textbf{GT}}
        \stackunder[2pt]{\includegraphics[scale=0.46]{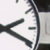}}{\textbf{$\times$2}}
        \stackunder[2pt]{\includegraphics[scale=0.46]{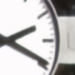}}{\textbf{$\times$3}}
        \stackunder[2pt]{\includegraphics[scale=0.46]{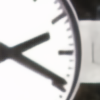}}{\textbf{$\times$4}}
        \stackunder[2pt]{\includegraphics[scale=0.46]{Figures/Multi_scale_SR_on_BurstSR/0105_0009_gt.png}}{\textbf{GT}}
        
        \vspace*{-6pt}
        \caption{\textbf{Visual comparison of multi-scale SR on \underline{BurstSR}}. BurstM predict several scaled-images with unimodel and the others are downsampling after $\times$4 SR.}
        \label{fig:multiscale_comparison_for_burstsr}
        \vspace{-20pt}
    \end{figure*}

    \noindent\textbf{SR on BurstSR~\cite{dbsr_burstsr_dataset}} Table \ref{tab:SR_result} is the quantitative comparison result of $\times4$ SR on BurstSR ~\cite{dbsr_burstsr_dataset}. Ours shows a performance gain of +0.17dB compared to the state-of-the-art baselines. As shown in Figure \ref{fig:x4_comparison_for_real_burst}, BurstM predicts outperform high-frequency details without the boundary artifacts from low-resolution images of real-world dataset. Also, BurstM predicts multi-scale SR without an additional processing, different with existing models. However, the quantitative comparison of $\times2$, $\times3$ is not possible, because there is no GT of both $\times2$, $\times3$ SR of given dataset~\cite{dbsr_burstsr_dataset}. We demonstrate the robustness of our approach by representing high-frequency details in Figure \ref{fig:multiscale_comparison_for_burstsr} and computation time in Table \ref{tab:Computation_parameter_timecheck}. BurstM directly predicts multiple scales, while other methods implement $\times4$ SR followed by bicubic downsampling. In Table \ref{tab:Computation_parameter_timecheck}, BurstM shows shorter computation time than the existing methods.
    \vspace{-20pt}

\begin{figure}[h]
    \footnotesize
    \centering
    \includegraphics[width=0.75in]{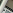}
    \includegraphics[width=0.75in]{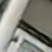}
    \includegraphics[width=0.75in]{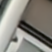}
    \includegraphics[width=0.75in]{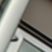}
    \includegraphics[width=0.75in]{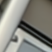}
    \includegraphics[width=0.75in]{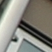}
    
    \includegraphics[width=0.75in]{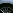}
    \includegraphics[width=0.75in]{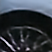}
    \includegraphics[width=0.75in]{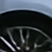}
    \includegraphics[width=0.75in]{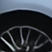}
    \includegraphics[width=0.75in]{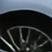}
    \includegraphics[width=0.75in]{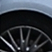}

    \stackunder[2pt]{\includegraphics[width=0.75in]{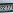}}{LR Input}
    \stackunder[2pt]{\includegraphics[width=0.75in]{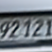}}{BIPNet}
    \stackunder[2pt]{\includegraphics[width=0.75in]{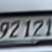}}{Burstormer}
    \stackunder[2pt]{\includegraphics[width=0.75in]{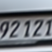}}{BSRT-L}
    \stackunder[2pt]{\includegraphics[width=0.75in]{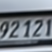}}{\textbf{BurstM}}
    \stackunder[2pt]{\includegraphics[width=0.75in]{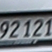}}{GT}
    
    \vspace*{-6pt}
    \caption{\textbf{Visual comparison for $\times$4 Super-Resolution for \underline{SynteticBurst}.}}
    \label{fig:x4_comparison_for_syntetic_burst}
    \vspace{-5pt}
    \end{figure}

\begin{figure}[h!]
    \centering
    \raisebox{0.2in}{\rotatebox{90}{$\times4$ SR}}
    \includegraphics[width=0.9in]{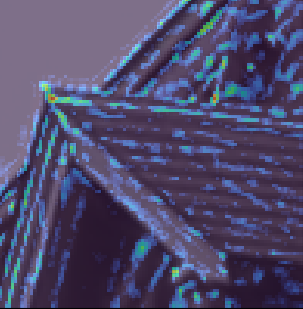}
    \includegraphics[width=0.9in]{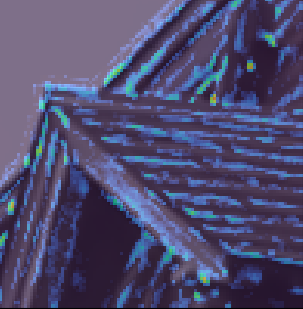}
    \includegraphics[width=0.9in]{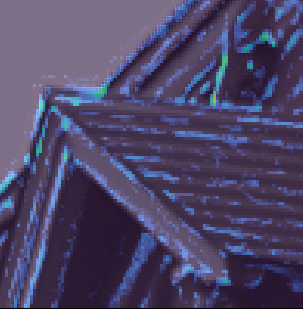}
    \includegraphics[width=0.9in]{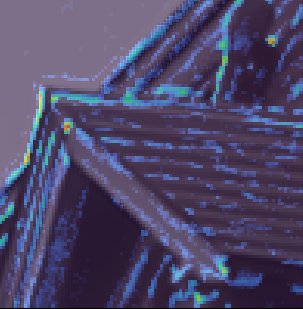}
    \includegraphics[width=0.9in]{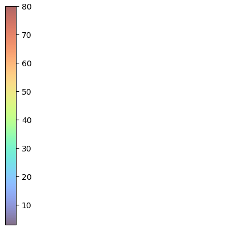}
    
    \raisebox{0.2in}{\rotatebox{90}{$\times3$ SR}}
    \includegraphics[width=0.9in]{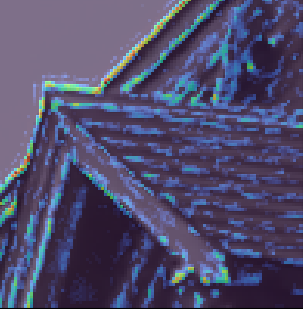}
    \includegraphics[width=0.9in]{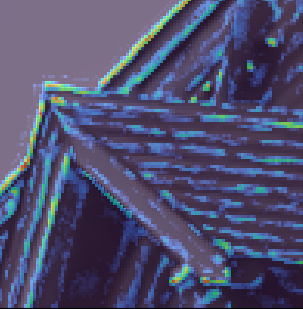}
    \includegraphics[width=0.9in]{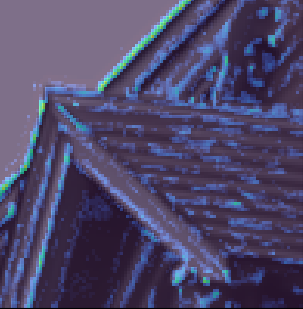}
    \includegraphics[width=0.9in]{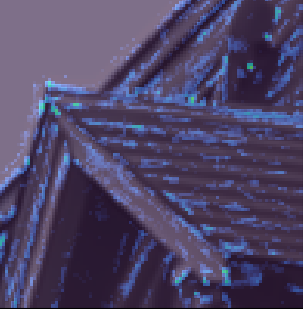}
    \includegraphics[width=0.9in]{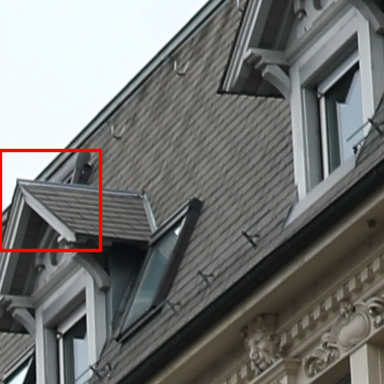}
    
    \raisebox{0.2in}{\rotatebox{90}{$\times2$ SR}}
    \stackunder[2pt]{\includegraphics[width=0.9in]{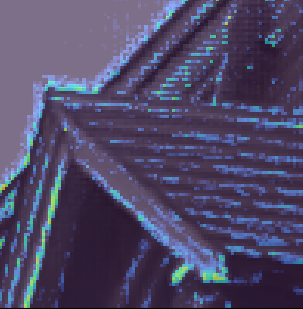}}{BIPNet}
    \stackunder[2pt]{\includegraphics[width=0.9in]{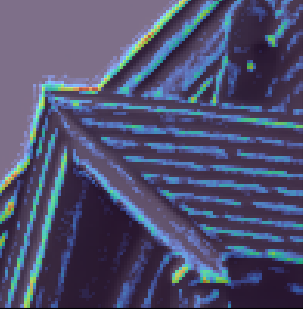}}{Burstormer}
    \stackunder[2pt]{\includegraphics[width=0.9in]{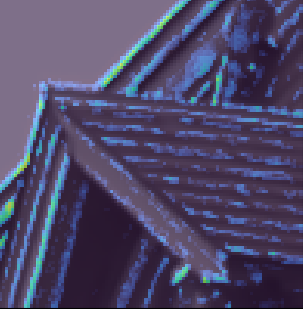}}{BSRT-L}
    \stackunder[2pt]{\includegraphics[width=0.9in]{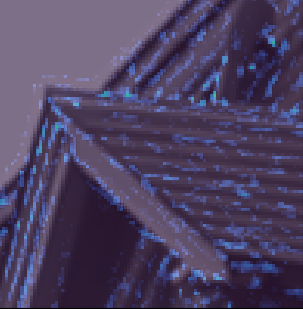}}{\textbf{BurstM}}
    \stackunder[2pt]{\includegraphics[width=0.9in]{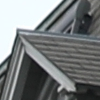}}{GT}
    \vspace*{-6pt}

        \caption{\textbf{Visual comparison for multi-scale SR on SyntheticBurst}. The color-bar indicates a difference with GT. The existing methods lose high-frequency textures during additional down-sampling because the highlighted locations are a high-frequency area. But BurstM represent the similar high-frequency textures with GT by because of direct prediction on multiple scale factors.}
        \label{fig:multiscale_comparison_for_synthetic}
        \vspace{-15pt}
    \end{figure}

    \noindent\textbf{SR on SyntheticBurst~\cite{synthetic_datset}} Table \ref{tab:SR_result} is the quantitative comparison results of multiple scale factors ($\times2$, $\times3$, $\times4$) SR and Figure \ref{fig:x4_comparison_for_syntetic_burst} is the visual comparison result of $\times4$ SR on SyntheticBurst~\cite{synthetic_datset}. Our model achieves second result with reasonable the number of parameters. In addition, we compare the performance on multiple scale factors in Table \ref{tab:SR_result} and Figure \ref{fig:multiscale_comparison_for_synthetic}. BurstM predicts high-resolution images on multiple scale factors with an unimodel and the others implement multiple SR with additional bicubic downsampling after $\times4$ SR. In case of BurstM, it is observed that PSNR gradually increases as the scale factor decrease. However, PSNR of existing methods decreases as the scale factor decreases because of the information loss during the additional bicubic downsampling process. Figure \ref{fig:multiscale_comparison_for_synthetic} shows the residual images, which means the loss of the high-frequency textures. Furthermore, as shown in Table \ref{tab:Computation_parameter_timecheck}, BurstM outperforms the other methods in terms of computation time.\\

\begin{figure}[h!]
    \centering
        \raisebox{0.2in}{\rotatebox{90}{IMAGE}}
        \includegraphics[width=0.8in]{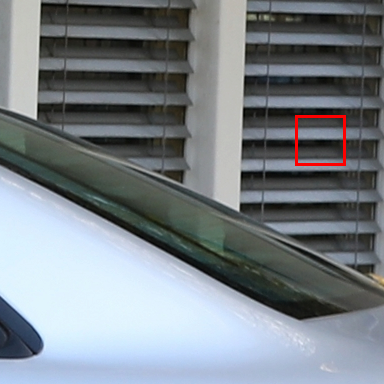}
        \includegraphics[width=0.8in]{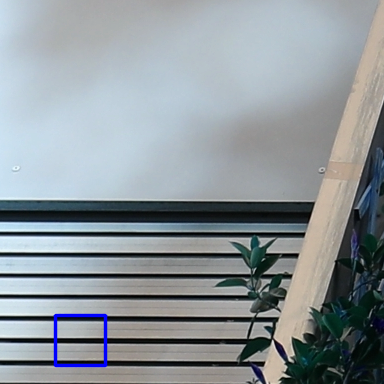}
        \includegraphics[width=0.8in]{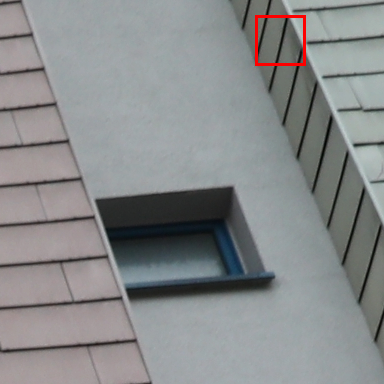}
        \includegraphics[width=0.8in]{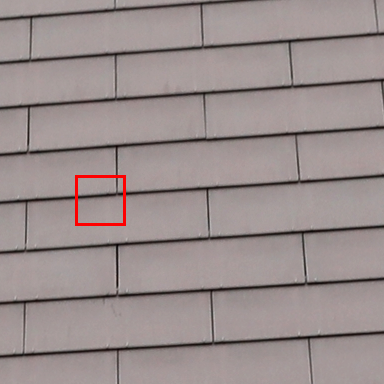}
        \includegraphics[width=0.8in]{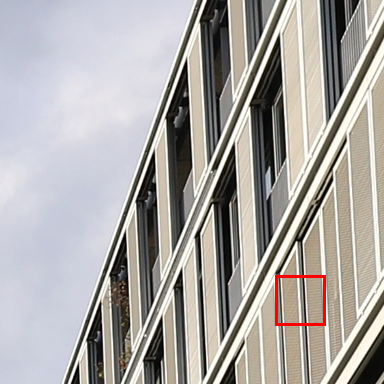}

        \raisebox{0.2in}{\rotatebox{90}{DFT}}
        \includegraphics[width=0.8in]{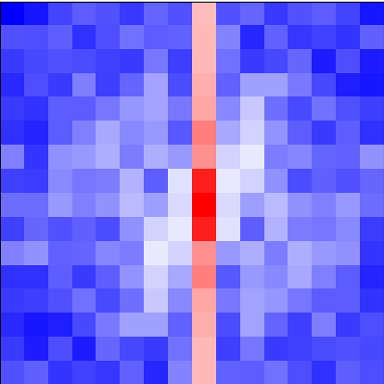}
        \includegraphics[width=0.8in]{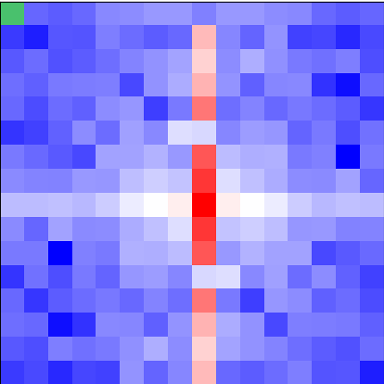}
        \includegraphics[width=0.8in]{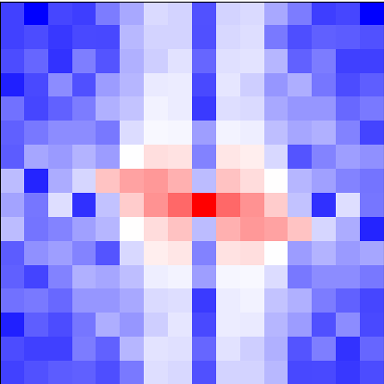}
        \includegraphics[width=0.8in]{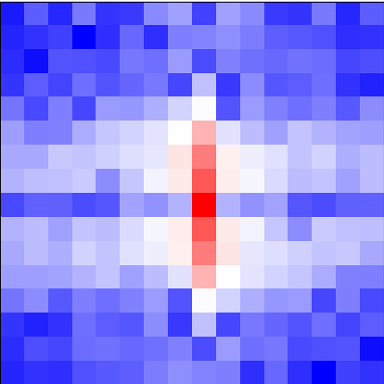}
        \includegraphics[width=0.8in]{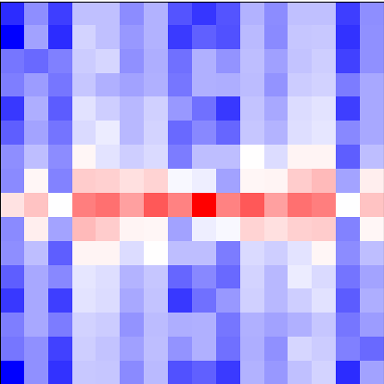}
        
        \raisebox{0.2in}{\rotatebox{90}{ACCUM.}}
        \includegraphics[width=0.8in]{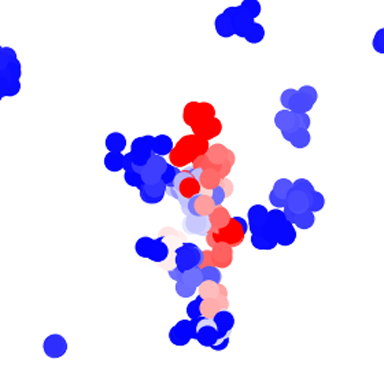}
        \includegraphics[width=0.8in]{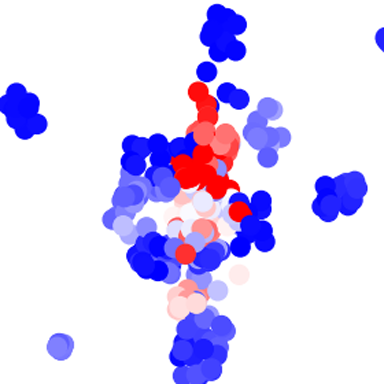}
        \includegraphics[width=0.8in]{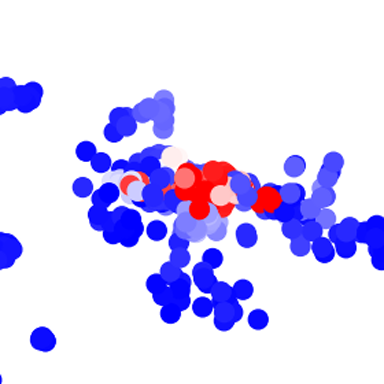}
        \includegraphics[width=0.8in]{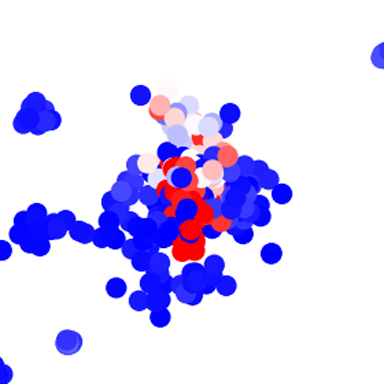}
        \includegraphics[width=0.8in]{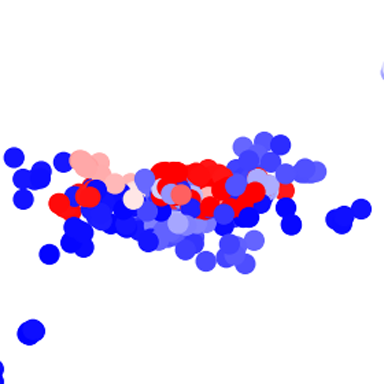}
        \vspace{-6pt}
    \caption{\textbf{Fourier space analysis.} Accumulated frequencies and coefficients from BurstM are analogous to DFT spectrum of GT (Red color in `DFT' and `ACCUM.').}
    \label{fig:Fourier_latent_space}
    \vspace*{-20pt}
    \end{figure}

    \noindent\textbf{Fourier feature space} Figure \ref{fig:Fourier_latent_space} demonstrates the estimation of high-frequency details on Fourier space. The result from each frame do not illustrate the DFT spectrum, but accumulated result of all frames shows a similar result with the DFT spectrum. Because BurstM leverages sub-pixel information from burst frames to represent the high-resolution images.

\begin{figure}[h!]
    \centering
    \begin{subfigure}[b]{1\textwidth}
        \centering
        \stackunder[2pt]{\includegraphics[width=2.0in, keepaspectratio]{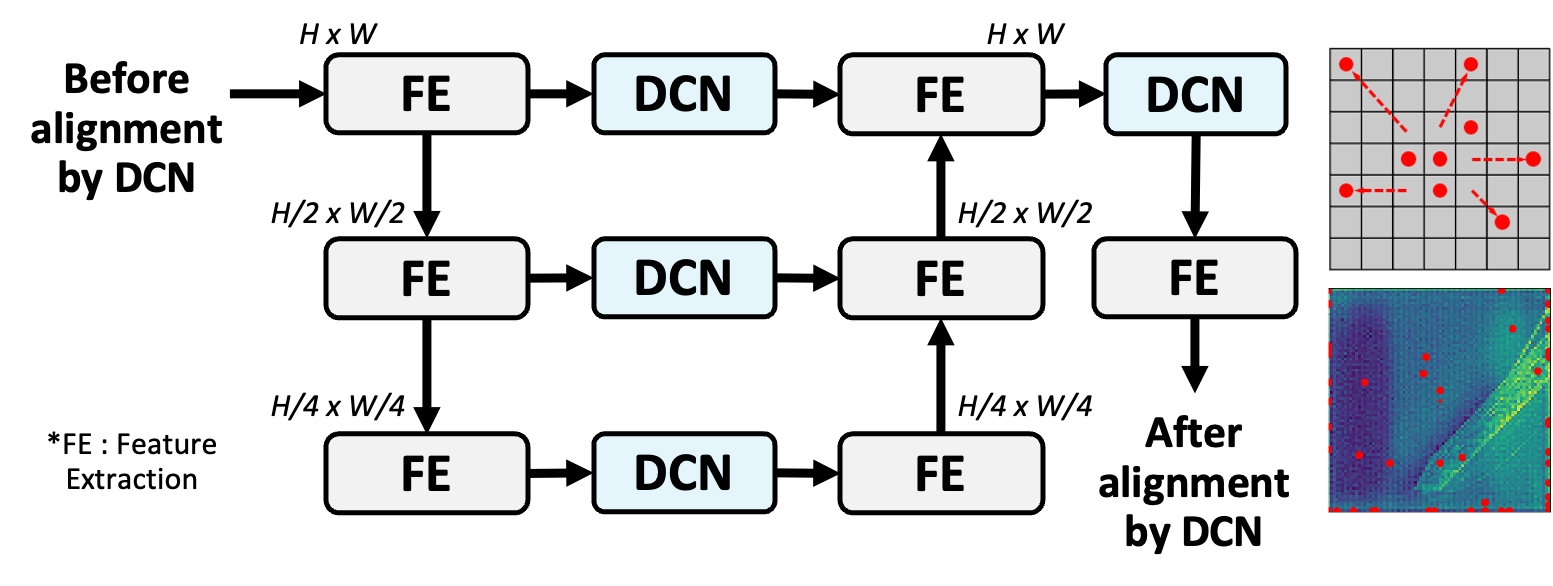}}{\textbf{DCN method flowchart~\cite{burstormer}}}
        \stackunder[2pt]{\includegraphics[width=2.0in, keepaspectratio]{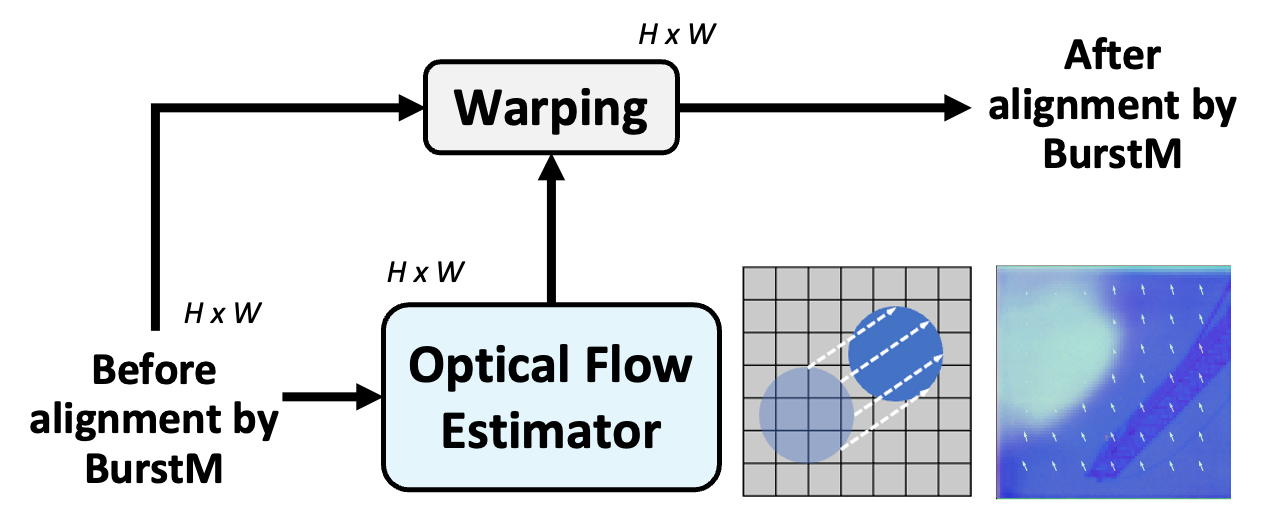}}{\textbf{BurstM method (\textit{ours}) flowchart}}
    \end{subfigure}\\
    \vspace{6pt}
    \vspace{-5pt}
    \begin{subfigure}[b]{0.95\textwidth}
      \raisebox{-4pt}{\rotatebox{90}{\scriptsize{Residual map}}}
      \raisebox{-4pt}{\rotatebox{90}{\scriptsize{Before align.}}}
      \includegraphics[width=0.68in,keepaspectratio]{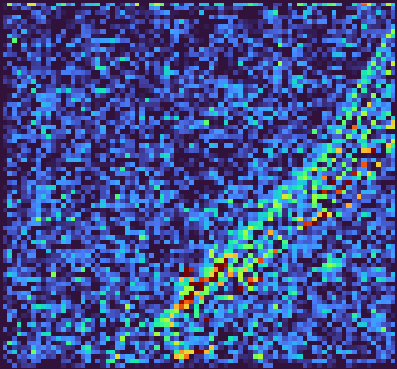}
      \includegraphics[width=0.68in,keepaspectratio]{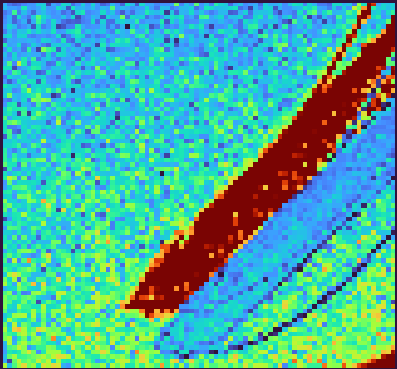}
      \includegraphics[width=0.68in,keepaspectratio]{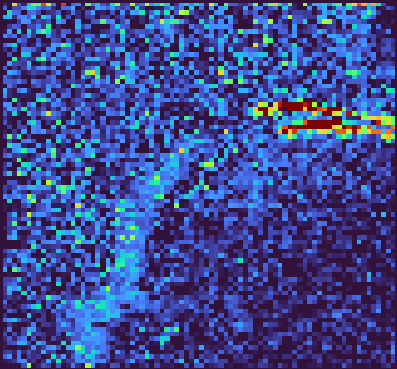}
      \includegraphics[width=0.68in,keepaspectratio]{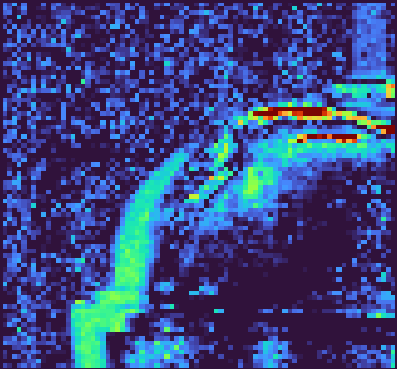}
      \includegraphics[width=0.68in,keepaspectratio]{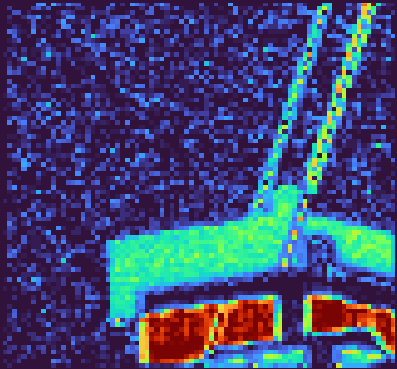}
      \includegraphics[width=0.68in,keepaspectratio]{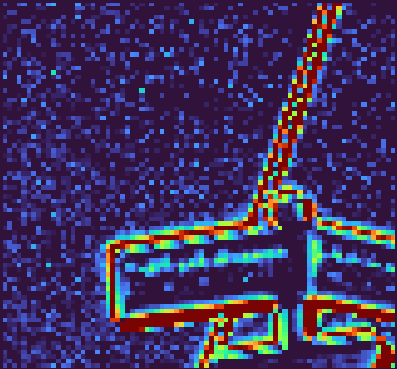}\\
      \raisebox{-4pt}{\rotatebox{90}{\scriptsize{Residual map}}}
      \raisebox{-4pt}{\rotatebox{90}{\scriptsize{After align.}}}
      \stackunder[2pt]
      {\includegraphics[width=0.68in,keepaspectratio]{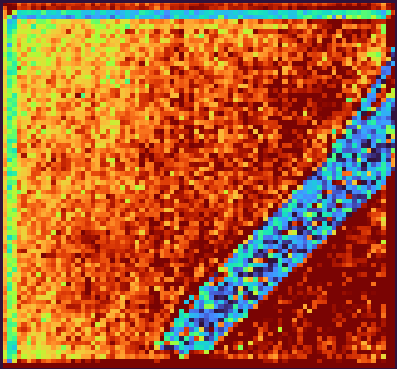}}{DCN}
      \stackunder[2pt]
      {\includegraphics[width=0.68in,keepaspectratio]{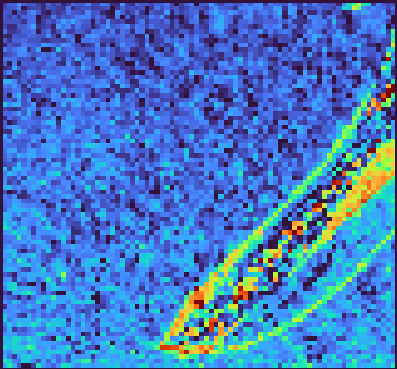}}{\textbf{Ours}}
      \stackunder[2pt]
      {\includegraphics[width=0.68in,keepaspectratio]{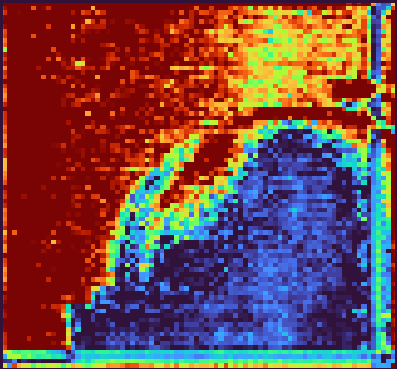}}{DCN}
      \stackunder[2pt]
      {\includegraphics[width=0.68in,keepaspectratio]{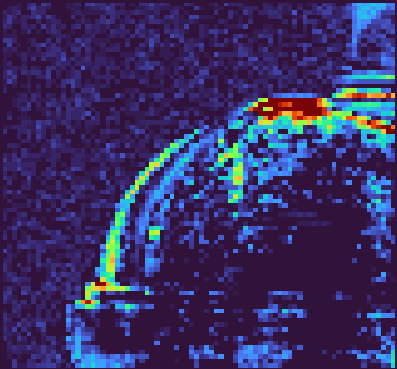}}{\textbf{Ours}}
      \stackunder[2pt]
      {\includegraphics[width=0.68in,keepaspectratio]{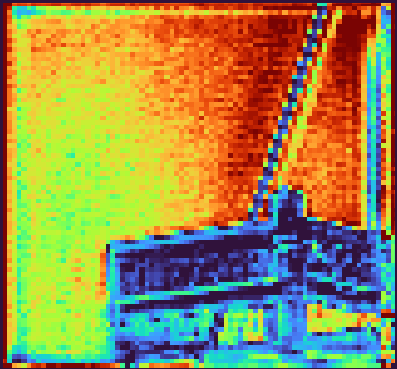}}{DCN}
      \stackunder[2pt]
      {\includegraphics[width=0.68in,keepaspectratio]{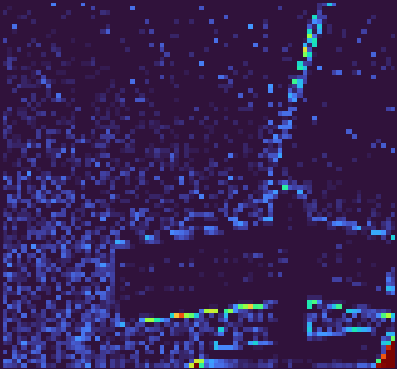}}{\textbf{Ours}}
    \end{subfigure}
    \hspace{-5pt}
    \begin{subfigure}[b]{0.03\textwidth}
        \includegraphics[height=1.35in,keepaspectratio]{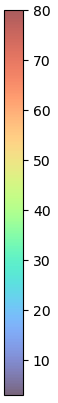}
    \end{subfigure}
    \vspace{-5pt}
    \caption{\textbf{Feature alignment results.} Flow charts indicate where we take features. The residual map is between the reference and source frames at each location in flowchart. DCN based-method has a big difference after an alignment because DCN take unrelated information during alignment. However, BurstM point out one-to-one correspondences between a reference and source frames. \underline{\textbf{BurstM shows accurate align result without the boundary artifacts.}}}
    \label{fig:Alignment_check}
    \vspace{-15pt}
    \end{figure}

    \noindent\textbf{Aligning performances} We conduct a comparative analysis of the alignment performance between DCN-based method and BurstM in Figure \ref{fig:Alignment_check}. As in~\cite{bipnet, gmtnet, burstormer, bsrt}, we utilize multi-layer DCN with varied spatial sizes for DCN-based method. In contrast, BurstM implements a single image warping process using predicted optical flows. The `Before align.' and `After align.' residual maps in Figure \ref{fig:Alignment_check} represent the differences with the reference frame at each position in flowchart. The red dots in `DCN method flowchart' illustrate offsets to the center pixel. DCN uses an unrelated pixel information, because DCN requires to use the number of predefined kernels regardless of the correlation. 
    Therefore, the aligned results are inaccurate and show high values in the residual map. Additionally, boundary artifacts are also observed due to the utilization of irrelevant information. The color map and white arrows in `BurstM method flowchart' are the visualization results of optical flow. BurstM method shows one-to-one correspondences between pixels. Thus, the alignment results demonstrate clearer and sharper compared to the DCN-based method without the boundary artifacts.\\
    \vspace{-30pt}

    \vspace{-10pt}

\section{Ablation study}
    \vspace{-0.1in}
    
    \noindent{\textbf{The comparison between with and without Fourier mapping}} INR~\cite{inr} is the most representative method for representing continuous signals, but it has a spectral bias~\cite{spectral_bias_one, spectral_bias_two, spectral_bias_three}. As shown in~\cite{lte, nis}, BurstM also addresses this issue through Fourier mapping. Table \ref{tab:Fourier_mapping} compares the performance of BurstM with and without Fourier mapping. BurstM without Fourier mapping has significantly lower PSNR values. This is because the network without Fourier mapping learns more about low frequencies due to spectral bias during training.\\
    
    \noindent{\textbf{Relationship between flexibility and performance}} BurstM is supporting the multiple scale SR ($\times2$, $\times3$, $\times4$). The performance in Table \ref{tab:SR_result} has trained with three multiple scale factors ($\times2$, $\times3$, $\times4$). We train with different number of scale factors such as one ($\times4$) and two ($\times3$, $\times4$) items. In such cases, we observed that PSNR increases on in-of-scale, but performance decreases on out-of-scale. When there are many instances of out-of-scale, the model's flexibility diminishes, so we proposed three scales satisfy both performance and flexibility.\\

    \noindent{\textbf{Different optical flow estimator}} While FNet~\cite{frvsr} estimates an optical flow based on given image sizes, the other optical flow estimator~\cite{ranjan2017optical} adjust image sizes according to its setting before input. This resizing of images introduces aliasing effects, resulting in a decline in SR performance. Additionally, BurstM with Spynet~\cite{ranjan2017optical} shows longer inference time than the proposed method. 
    \vspace{-0.3in}
    
    \begin{table}[h]
        \begin{subtable}[h]{\columnwidth}
            \centering
            \setlength{\tabcolsep}{1.5pt}
            \begin{tabular}{c|c|c}
                & BurstM \textbf{w\textbackslash o Fourier mapping} & BurstM \textbf{w\textbackslash Fourier mapping} \\
                \hline
                $\times4$ PSNR & 39.82 dB & \textcolor{red}{41.74 dB}
                \end{tabular}
                \subcaption{The performance comparison between with and without Fourier mapping}
                \vspace{2pt}
                \label{tab:Fourier_mapping}
        \end{subtable}
        \vspace{-0.1in}
        \\
        \begin{subtable}[h]{0.5\columnwidth}
            \centering
            \setlength{\tabcolsep}{1.5pt}
            \begin{tabular}{c|ccc}
                Training scale              & $\times2$ & $\times3$ & $\times4$\\\hline\hline
                $\times4$                   &  33.36 & 35.97 & \textcolor{red}{42.26} \\
                $\times3, \times4$          &  30.80 & \textcolor{red}{43.61} & \textcolor{red}{41.82} \\
                $\times2, \times3, \times4$ (Ours) &  \textcolor{red}{44.77} & \textcolor{red}{43.45} & \textcolor{red}{41.74} 
                
            \end{tabular}
            \captionsetup{justification=centering}
            \subcaption{Relationship between flexibility and performance (PSNR:dB)}
            \label{tab:scale_config}
            \vspace{-5pt}
        \end{subtable}
        \hspace{-0.15in}
        \begin{subtable}[h]{0.5\columnwidth}
            \centering
            \setlength{\tabcolsep}{1.5pt}
            \begin{tabular}{c|c|c}
                 & \begin{tabular}[c]{@{}c@{}}BurstM\\ w\textbackslash \textbf{Spynet}\end{tabular} & \begin{tabular}[c]{@{}c@{}}BurstM\\ w\textbackslash \textbf{FNet} (Ours) \end{tabular}\\ \hline\hline
            $\times4$ PSNR & 40.85 dB & \textcolor{red}{41.74 dB}\\
            Inference time & 14.3 ms  & \textcolor{red}{11.6 ms} \\
            \end{tabular}
            \captionsetup{justification=centering}
            \subcaption{Performance comparison \\based on optical flow estimator}
            \label{tab:opticalflow_config}
            \vspace{-5pt}
        \end{subtable}
    
        \caption{\textbf{Ablation study summary}\\
        (a) Fourier mapping contributes to the model performance.\\
        (b) \textcolor{red}{Red} color indicates the result of in-scale. Less training scale has higher PSNR on in-scale but network flexibility is low due to increasing out-scale. Thus, we select our model with 3 scale factors ($\times2,\times3,\times4$) considering both flexibility and performance.\\
        (c) A different optical flow estimator leads to PSNR drop and an increase in inference time.}
        \vspace{-0.4in}
    \end{table}

\vspace{-0.2in}
\section{Discussion}
\begin{figure}[h!]
        \centering
        \begin{subfigure}[b]{0.8\textwidth}
            \stackunder[0pt]{    
                \includegraphics[width=0.68in,keepaspectratio]{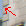}
                \includegraphics[width=0.68in,keepaspectratio]{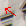}
                \includegraphics[width=0.68in,keepaspectratio]{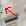}
                \includegraphics[width=0.68in,keepaspectratio]{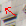}
                \includegraphics[width=0.68in,keepaspectratio]{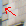}}{}
            \stackunder[0pt]{    
                \includegraphics[width=0.68in,keepaspectratio]{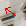}
                \includegraphics[width=0.68in,keepaspectratio]{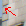}
                \includegraphics[width=0.68in,keepaspectratio]{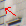}
                \includegraphics[width=0.68in,keepaspectratio]{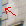}
                \includegraphics[width=0.68in,keepaspectratio]{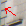}}{}
                
            \stackunder[2pt]{
                \includegraphics[width=0.68in,keepaspectratio]{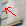}
                \includegraphics[width=0.68in,keepaspectratio]{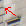}
                \includegraphics[width=0.68in,keepaspectratio]{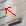}
                \includegraphics[width=0.68in,keepaspectratio]{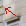}
                \includegraphics[width=0.68in,keepaspectratio]{Figures/white.png}}{Input frames}
        \end{subfigure}
        \hspace{-15pt}
        \begin{subfigure}[b]{0.2\textwidth}
          \stackunder[0pt]{
              \raisebox{0.2in}{\rotatebox{90}{\scriptsize{GT}}}
              \includegraphics[width=0.68in,keepaspectratio]{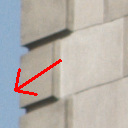}}{}
          \stackunder[0pt]{    
              \raisebox{0.15in}{\rotatebox{90}{\scriptsize{BSRT-L}}}
              \includegraphics[width=0.68in,keepaspectratio]{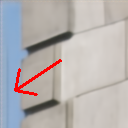}}{}
              
          \stackunder[2pt]{
              \raisebox{0.15in}{\rotatebox{90}{\scriptsize{\textbf{BurstM}}}}
              \includegraphics[width=0.68in,keepaspectratio]{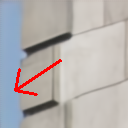}}{Result}
        \end{subfigure}
        \vspace{-5pt}
        \caption{\textbf{Boundary artifacts} The most of frames didn't provide the information of edge. The existing method~\cite{bsrt} suffers the boundary line artifacts, but \textbf{BurstM represents clear images} \underline{\textbf{without the boundary line artifacts.}}}
        \label{fig:Boundary_artfifact}
        \vspace{-0.3in}
    \end{figure}

    \vspace{-0.1in}
     \noindent{\textbf{Boundary line artifacts}} As shown in Figure \ref{fig:Boundary_artfifact}, the recent method~\cite{bsrt} exhibits boundary artifacts despite utilizing optical flow, due to its continued reliance on DCN~\cite{deformable_convolution_network} for alignment. As illustrated in Figure \ref{fig:Alignment_check} and \ref{fig:Boundary_artfifact}, DCN~\cite{deformable_convolution_network} must utilize information corresponding to the predefined kernel size regardless of correlation, resulting in the incorporation of unnecessary information when most frames fail to provide essential data. In contrast, BurstM aligns frames without DCN~\cite{deformable_convolution_network}, solely utilizing optical flow. Consequently, BurstM avoids the use of irrelevant information, leading to accurate predictions without boundary artifacts.\\
     \vspace{-0.35in}

     \noindent{\textbf{Computation complexity}} Despite advancements in computational system, SR field still requires shorter computation times as the image resolutions increases. Table \ref{tab:Computation_parameter_timecheck} compares the computation time usage of BurstM with the other methods~\cite{bipnet, burstormer, bsrt}. We implements $\times4$ SR with $80\times80$ frame sizes of BurstSR dataset~\cite{dbsr_burstsr_dataset} on NVIDIA RTX3090 24GB. While BurstM has a large number of parameters, BurstM demonstrates the shortest computation time.
     \vspace{-0.3in}
    
    \begin{table}[h]
        \begin{minipage}{0.57\linewidth}
            \includegraphics[width=\linewidth,keepaspectratio]{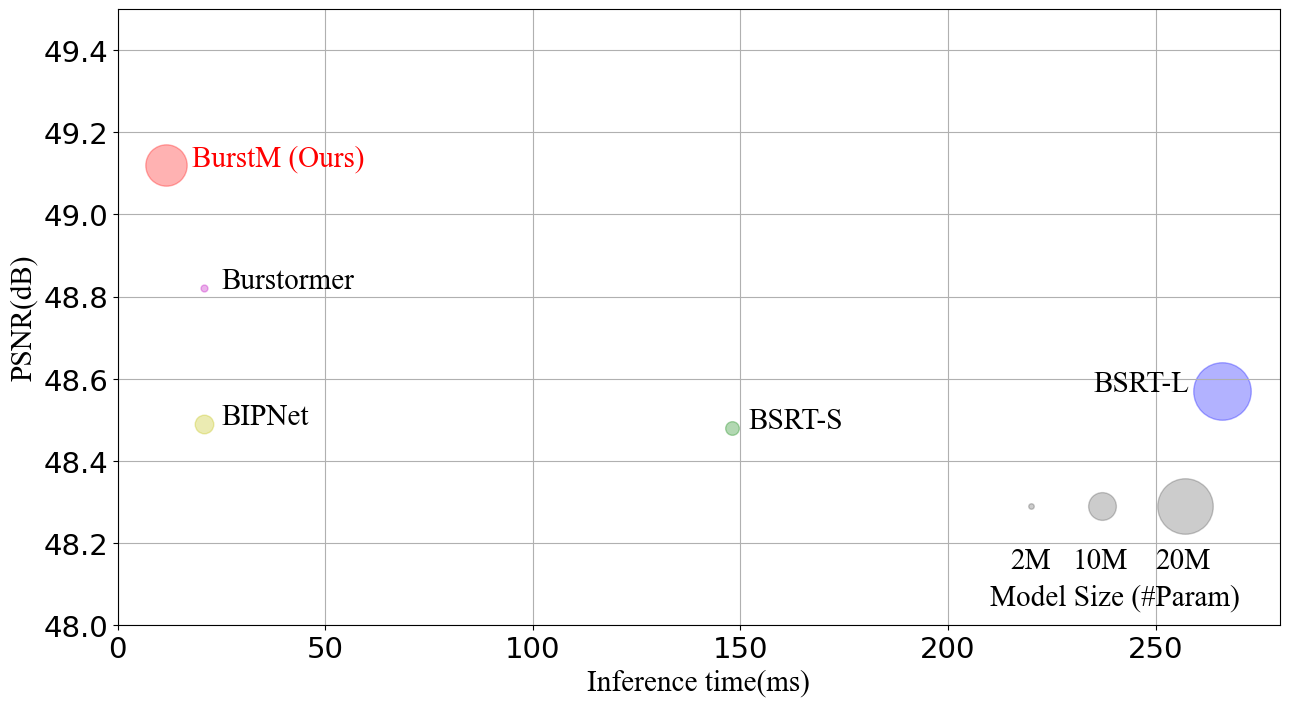}
        \end{minipage}
        \hspace{-0.08in}
        \begin{minipage}{0.4\linewidth}
            \scriptsize
            \begin{tabular}{c|c|c|c}
            \multicolumn{1}{c|}{\begin{tabular}[c|]{@{}c@{}}SR scale: \\ $\times4$\end{tabular}} & \multicolumn{1}{c|}{\begin{tabular}[c|]{@{}c@{}}Params.(M)\\ / Mem.(GB)\end{tabular}} & \multicolumn{1}{c|}{\begin{tabular}[c|]{@{}c@{}}Inference\\ time(ms)\end{tabular}} & \multicolumn{1}{c}{\begin{tabular}[c]{@{}c@{}}PSNR
            \\ (dB)\end{tabular}} 
             \\ \hline
            \multicolumn{1}{c|}{\begin{tabular}[c|]{@{}c@{}}BIPNet$^{\dag}$\\ \cite{bipnet}\end{tabular}} & \multicolumn{1}{c|}{6.7 / 11.8 }  & \multicolumn{1}{c|}{20.8} & \multicolumn{1}{c}{48.49} \\
            \multicolumn{1}{c|}{\begin{tabular}[c|]{@{}c@{}}Burstormer$^{\dag}$\\ \cite{burstormer}\end{tabular}} & \multicolumn{1}{c|}{\textcolor{red}{2.5} / 12.0} & \multicolumn{1}{c|}{20.8} & \multicolumn{1}{c}{48.82} \\
            \multicolumn{1}{c|}{\begin{tabular}[c|]{@{}c@{}}BSRT-S$^{\dag}$\\ \cite{bsrt}\end{tabular}} & \multicolumn{1}{c|}{4.92 / \textcolor{red}{1.0}} & \multicolumn{1}{c|}{148.0} & \multicolumn{1}{c}{48.48}  \\
            \multicolumn{1}{c|}{\begin{tabular}[c|]{@{}c@{}}BSRT-L$^{\dag}$\\ \cite{bsrt}\end{tabular}} & \multicolumn{1}{c|}{20.71 / 1.5} & \multicolumn{1}{c|}{266.1} & \multicolumn{1}{c}{48.57}  \\ \hline
            \multicolumn{1}{c|}{\begin{tabular}[c|]{@{}c@{}}\textbf{BurstM}\\ \textbf{(Ours)}\end{tabular}} & \multicolumn{1}{c|}{14.9 / 8.4} & \multicolumn{1}{c|}{\textcolor{red}{11.6}} & \multicolumn{1}{c}{\textcolor{red}{49.12}}                                      
            \end{tabular}
        \end{minipage}

    \caption{\textbf{Computational complexity on one GPU (RTX3090).} Despite a large number of parameters, BurstM shows the inference time (ms) and best PSNR (dB) performance. \textcolor{red}{Red} : best metrics. ${\dag}$ indicates Transformer-based method.}
    \label{tab:Computation_parameter_timecheck}
    \vspace*{-0.3in}
    \end{table}
    \noindent{\textbf{Super-Resolution with floating scale factors}} BurstM can estimate the multi-scale SR with floating scale factors such as $\times 3.5$ without any architecture changes. However, the training of the floating scale factors ($\times2$$\sim$$\times4$) shows lower PSNR (39.44dB) than the training of the integer scale factors (41.74dB). The performance drop is caused by the demosaic process when generating low-resolution frames. To convert from `sRGB' to `RGGB', the dimensions (H, W) of the images are changed to (H/2, W/2). If the image size of `sRGB' is not divisible by two when a floating scale factor is applied, the network apply an additional downsampling operation to make the dimension of (H/2, W/2). The double downsampling imposes a severe degradation of the generated low-resolution images, and it results in low PSNRs.\\

\begin{figure}[h]
    \footnotesize
    \vspace*{-0.3in}
    \centering
    \raisebox{0.1in}{\rotatebox{90}{w/o align.}}
    \includegraphics[width=0.8in]{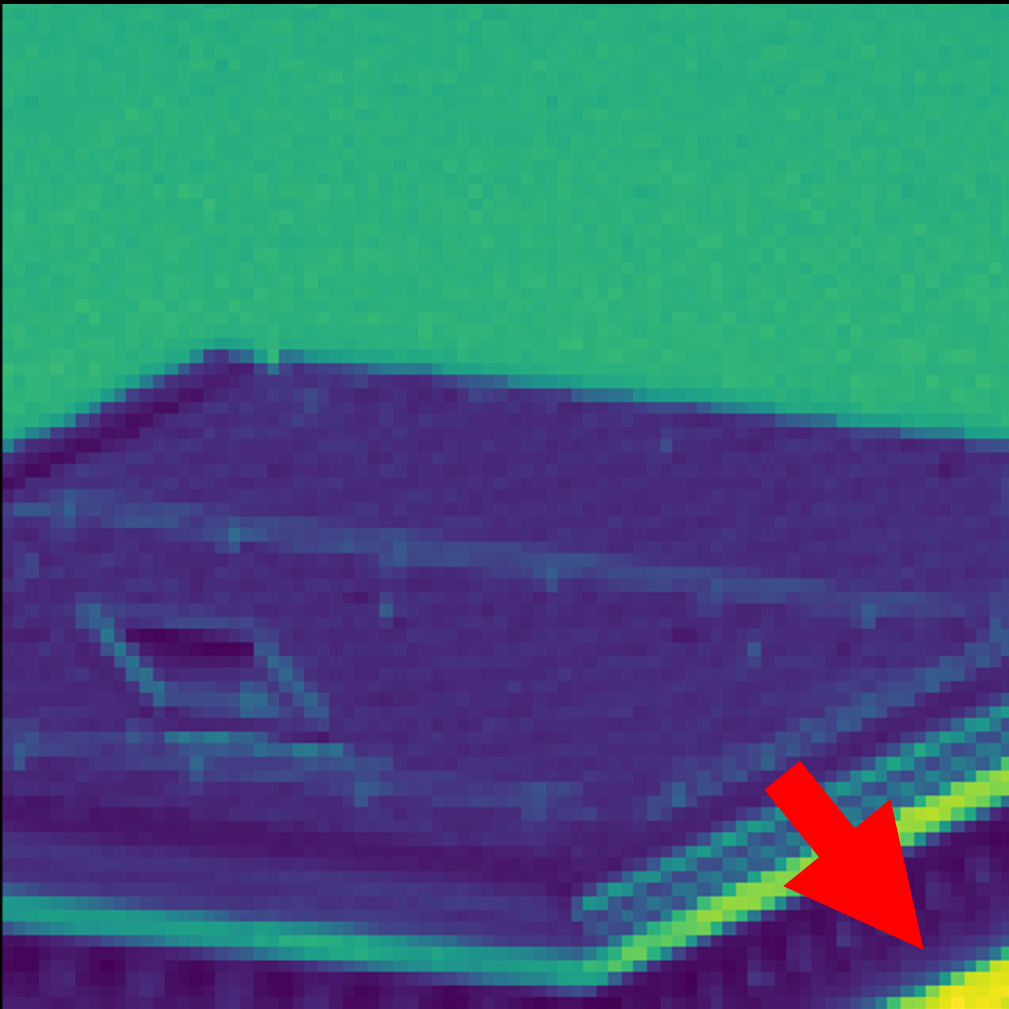}
    \includegraphics[width=0.8in]{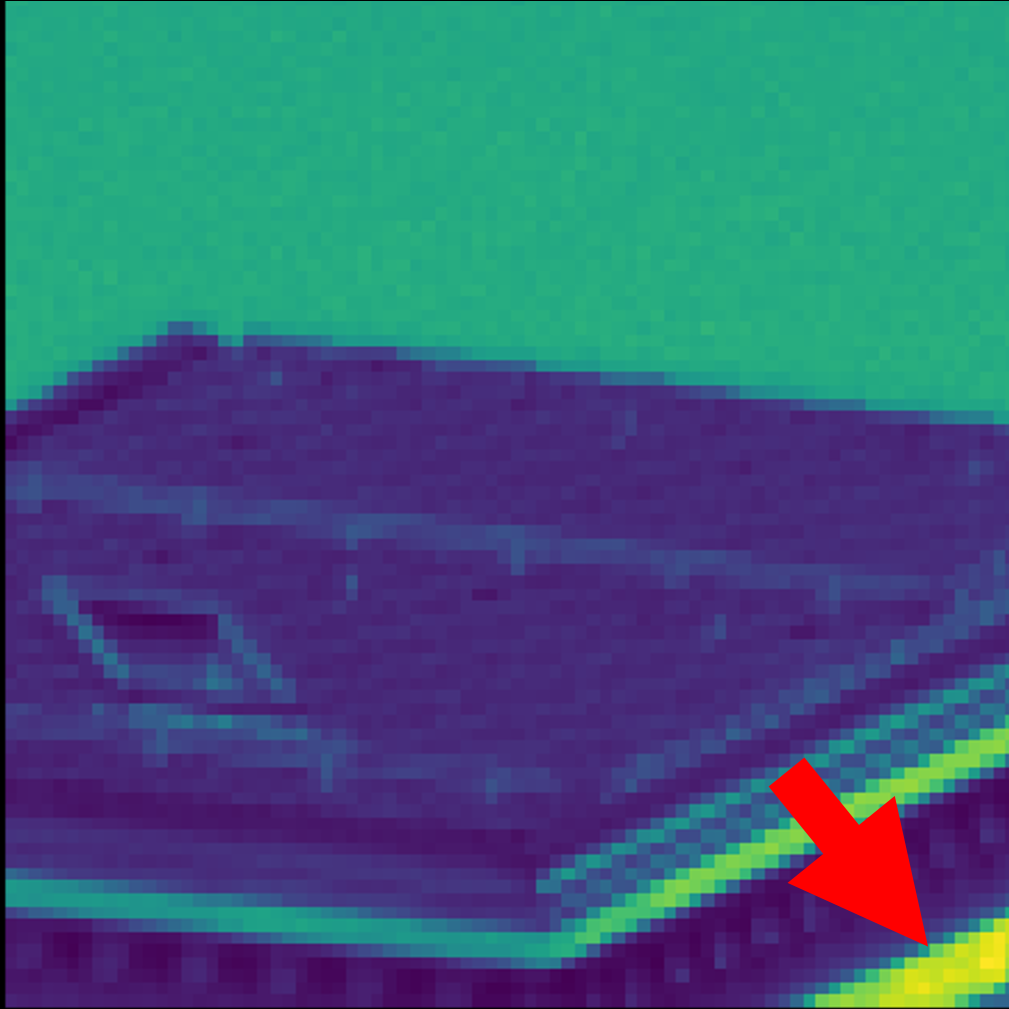}
    \includegraphics[width=0.8in]{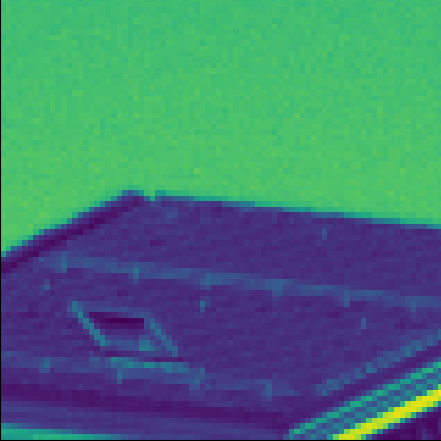}
    \includegraphics[width=0.8in]{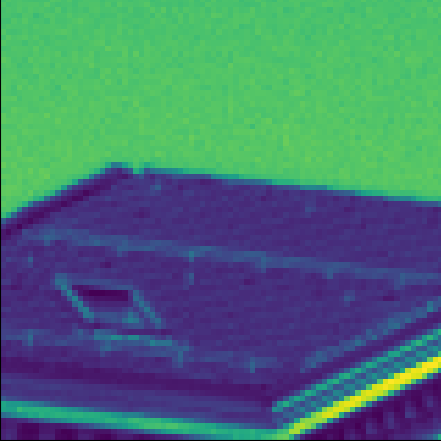}
    \includegraphics[width=0.8in]{Figures/white.png}
    
    \raisebox{0.01in}{\rotatebox{90}{Optical flow}}
    \includegraphics[width=0.8in]{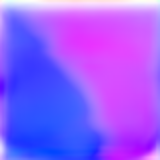}
    \includegraphics[width=0.8in]{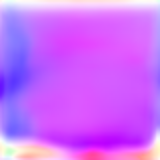}
    \includegraphics[width=0.8in]{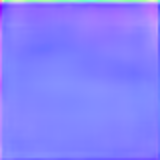}
    \includegraphics[width=0.8in]{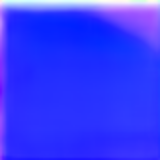}
    \includegraphics[width=0.8in]{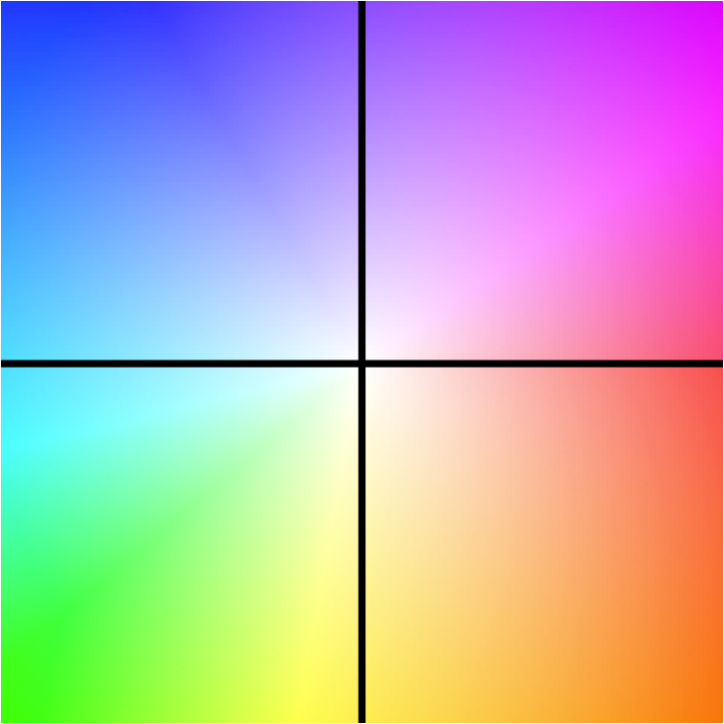}
    
    \raisebox{0.1in}{\rotatebox{90}{w/ align.}}
    \stackunder[2pt]{\includegraphics[width=0.8in]{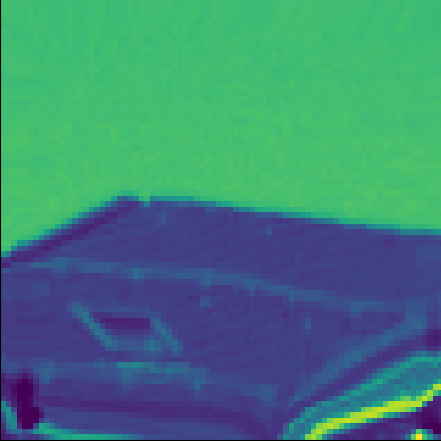}}{Corner case}
    \stackunder[2pt]{\includegraphics[width=0.8in]{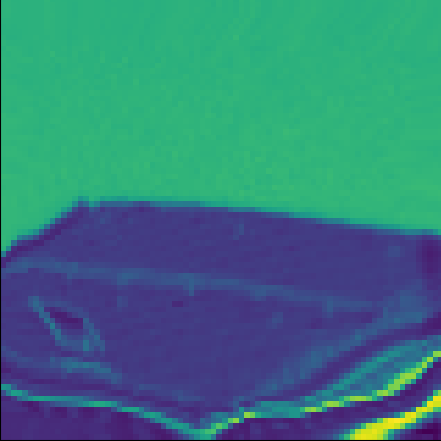}}{Corner case}
    \stackunder[2pt]{\includegraphics[width=0.8in]{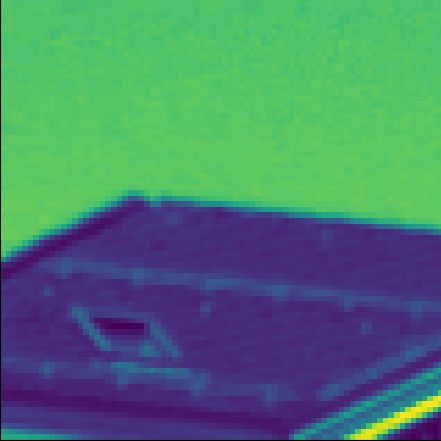}}{Normal case}
    \stackunder[2pt]{\includegraphics[width=0.8in]{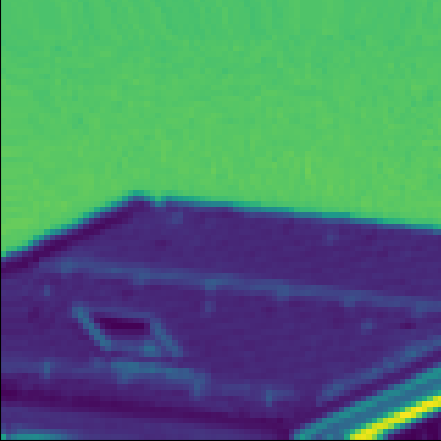}}{Normal case}
    \stackunder[2pt]{\includegraphics[width=0.8in]{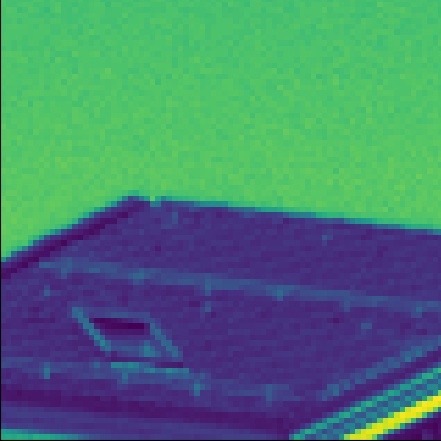}}{Reference}
    
    \vspace*{-6pt}
        \caption{\textbf{Corner case.} Inaccurate optical flow prediction due to occlusions which are caused by new objects (red arrow) not present in the reference image appear.}
        \label{fig:corner_case}
    \vspace*{-0.5in}
    \end{figure}

    \vspace*{-0.1in}
    \noindent{\textbf{Corner case}} The major challenge on optical flow is occlusions which are caused by a new object that is not present in the reference frame. BurstM also has the occlusion issue (Figure \ref{fig:corner_case}) even though BurstM demostrate accurate alignment performance than DCN based methods. Fortunately, BurstM utilize inaccurate optical flow of an occlusion as meaningful prior and reconstruct the high-quality images in a high-resolution grid.
    
\section{Conclusion}
    \vspace{-0.1in}
    In this paper, we proposed Deep Burst Multi-scale SR using Fourier Space with Optical Flow (BurstM) to overcome the limitations of existing methods: SR of fixed scale factor ($\times4$) and low-quality of real-world dataset~\cite{dbsr_burstsr_dataset}. Our network performs the precise alignment using optical flow and the advanced warping process based on estimated Fourier information. We demonstrated their effectiveness in representing the high-frequency textures with +0.17dB PSNR gain in the real-world dataset~\cite{dbsr_burstsr_dataset} compared to the previous state-of-the-art model. Furthermore, BurstM successfully implements multi-scale SR ($\times2, \times3, \times4$) of the unimodel by representing high-frequency details using the INR method. The function of multi-scale SR enhances our model's capacity compared to the other learning-based methods. In addition, we demonstrate our approach is faster inference time compared to the existing methods by the efficient computation.
    
    \textbf{Acknowledgement} This work was supported by Institute of Information \& communications Technology Planning \& Evaluation (IITP) grant funded by the Korea government(MSIT) (RS-2021-II212068, Artificial Intelligence Innovation Hub) and the National Research Foundation of Korea(NRF) grant funded by the Korea government(MSIT) (RS-2024-00335741).



%
%
\bibliographystyle{splncs04}
\bibliography{main}
\end{document}